\documentclass[sigconf]{acmart}

\usepackage{graphicx}
\usepackage{balance}  %

\usepackage{graphics}
\usepackage{mdwlist}
\usepackage[ruled, noend, linesnumbered]{algorithm2e} %

\usepackage{booktabs}
\usepackage{soul}
\usepackage[center]{subfigure}

\usepackage{amsmath}
\usepackage{amsfonts}
\usepackage{amsbsy}
\usepackage{amsthm}
\usepackage[]{caption}
\usepackage{multirow}
\usepackage{mathrsfs}
\usepackage[mathscr]{eucal} %
\usepackage{xspace}
\usepackage{paralist} %
\usepackage{mathtools}
\usepackage{verbatim}
\usepackage{tabularx}
\usepackage{makecell}

\usepackage{enumitem}
\usepackage{pifont}%

\newtheorem{problem}{Problem}
\newtheorem{lemma}{Lemma}
\newtheorem{theorem}{Theorem}
\newtheorem{corollary}{Collorary}
\newtheorem*{proof}{Proof}

\newcommand{\smallsection}[1]{{\vspace{0.3mm}\noindent {\bf{\underline{\smash{#1}}}}}}

\definecolor{mygreen}{RGB}{0, 100, 0}
\newcommand{\cmark}{\textcolor{mygreen}{\ding{51}}}%
\newcommand{\xmark}{\textcolor{red}{\ding{55}}}%
\let\oldnl\nl%
\newcommand{\nonl}{\renewcommand{\nl}{\let\nl\oldnl}}%
\newenvironment{proofof}[1]{\noindent{\bf Proof of #1.}}%
{\hspace*{\fill}$\Box$\par\vspace{4mm}}

\newcommand{\greedy}{\textsc{Greedy}\xspace}
\newcommand{\sags}{\textsc{SAGS}\xspace}
\newcommand{\sweg}{\textsc{SW}e\textsc{G}\xspace}
\newcommand{\rnd}{\textsc{Randomized}\xspace}

\newcommand{\sgreedy}{\textsc{MoSSo-{Greedy}}\xspace}
\newcommand{\mcmc}{\textsc{MoSSo-{MCMC}}\xspace}
\newcommand{\algobasic}{\textsc{MoSSo-Simple}\xspace}
\newcommand{\getrnd}{\textsc{getRandomNeighbor}\xspace}
\newcommand{\algo}{\textsc{MoSSo}\xspace}

\newcommand{\GISTAR}{G_t^*}
\newcommand{\GIPSTAR}{G_{t+1}^*}
\newcommand{\CI}{C_{t}}
\newcommand{\CIP}{C_{t+1}}
\newcommand{\summary}{(S, P)}
\newcommand{\summaryt}{(S_t, P_t)}

\newcommand{\Cp}{C^+}
\newcommand{\Cm}{C^-}
\newcommand{\cor}{(C^+, C^-)}
\newcommand{\cort}{(C^+_t, C^-_t)}
\newcommand{\stream}{\{e_t\}_{t=0}^{\infty}}
\newcommand{\edge}{\{u, v\}}
\newcommand{\ea}{\{u, v\}^+}
\newcommand{\ed}{\{u, v\}^-}
\newcommand{\escape}{e}

\newcommand{\alone}{\textsc{Corrective Escape}\xspace}
\newcommand{\fastr}{\textsc{Fast Random}\xspace}
\newcommand{\guide}{\textsc{Careful Selection}\xspace}

\newcommand{\tn}{testing node\xspace}
\newcommand{\tp}{testing pool\xspace}
\newcommand{\cp}{candidate pool\xspace}
\newcommand{\cand}{candidate\xspace}

\newcommand{\target}{y}
\newcommand{\tnsymb}{TN(u)}
\newcommand{\tpsymb}{TP(u)}
\newcommand{\cpsymb}{CP(y)}

\newcommand{\candidate}{z}
\newcommand{\candidateS}{S_z}

\newcommand{\fprob}{\textsc{Obstructive Obsession}\xspace}
\newcommand{\sprob}{\textsc{Costly Neighborhood Retrievals}\xspace}
\newcommand{\tprob}{\textsc{Redundant Tests}\xspace}

\newcommand{\ratio}{compression ratio\xspace}
\newcommand{\ratios}{compression ratios\xspace}
\renewenvironment{proof}{\textsc{Proof.}}{\hfill$\blacksquare$}

\copyrightyear{2020}
\acmYear{2020}
\setcopyright{acmcopyright}\acmConference[KDD '20]{Proceedings of the 26th ACM SIGKDD Conference on Knowledge Discovery and Data Mining}{August 23--27, 2020}{Virtual Event, USA}
\acmBooktitle{Proceedings of the 26th ACM SIGKDD Conference on Knowledge Discovery and Data Mining (KDD '20), August 23--27, 2020, Virtual Event, USA}
\acmPrice{15.00}
\acmDOI{10.1145/3394486.3403074}
\acmISBN{978-1-4503-7998-4/20/08}

\begin{document}
\fancyhead{}
\title{Incremental Lossless Graph Summarization}

\author{Jihoon Ko}
\authornote{Equal Contribution. \footnotemark[2]Corresponding author.}
\affiliation{%
\institution{KAIST AI}
}
\email{jihoonko@kaist.ac.kr}

\author{Yunbum Kook}
\authornotemark[1]
\affiliation{%
\institution{Dept. of Mathematical Science, KAIST}
}
\email{yb.kook@kaist.ac.kr}

\author{Kijung Shin}
\authornotemark[2]
\affiliation{%
\institution{KAIST AI \& EE}
}
\email{kijungs@kaist.ac.kr}

\begin{abstract}
	\vspace{-1mm}
	{\it Given a fully dynamic graph, represented as a stream of edge insertions and deletions, how can we obtain and incrementally update a lossless summary of its current snapshot?}

As large-scale graphs are prevalent, concisely representing them is inevitable for efficient storage and analysis.
Lossless graph summarization is an effective graph-compression technique with many desirable properties. It aims to compactly represent the input graph as %
(a) a \textit{summary graph} consisting of supernodes (i.e., sets of nodes) and superedges (i.e., edges between supernodes), which provide a rough description, and (b) \textit{edge corrections} which fix errors induced by the rough description.
While a number of batch algorithms, suited for static graphs, have been developed for rapid and compact graph summarization, they are highly inefficient in terms of time and space for dynamic graphs, which are common in practice.

In this work, we propose \algo, the first incremental algorithm for lossless summarization of fully dynamic graphs.  
In response to each change in the input graph, 
\algo updates the output representation by repeatedly moving nodes among supernodes. 
\algo decides nodes to be moved and their destinations carefully but rapidly based on several novel ideas.
Through extensive experiments on 10 real graphs, we show \algo is 
(a) \textbf{Fast and `any time'}: processing each change in near-constant time (less than $0.1$ millisecond), up to {\bf $\textbf{7}$ orders of magnitude faster} than running state-of-the-art batch methods,
(b) \textbf{Scalable}: summarizing graphs with hundreds of millions of edges, requiring sub-linear memory during the process, and
(c) \textbf{Effective}: achieving comparable compression ratios even to state-of-the-art batch methods.

	\vspace{-1mm}
\end{abstract}

\maketitle

\vspace{-2mm}
\section{Introduction}
\vspace{-1mm}
\label{sec:intro}
\begin{table}[t]
	\centering
	\small
	\caption{\label{tab:compare} Comparison of lossless graph summarization methods. \algo is fast, space-efficient, and online.}
	\scalebox{0.9}{
	\begin{tabular}{l||ccc|c}
		\toprule
		& \cite{navlakha2008graph} & \cite{khan2015set} & \cite{shin2019sweg} & \algo (Proposed) \\
		\midrule
		Takes near-linear time           &  \xmark          & \cmark    & \cmark &    \cmark               \\
		Requires sub-linear space            & \xmark          & \xmark    & \xmark &      \cmark             \\
		Handles inserted edges (or nodes)          & \xmark          & \xmark    & \xmark        &      \cmark             \\
		Handles deleted edges (or nodes)      & \xmark          & \xmark    & \xmark        &      \cmark             \\ 
		\bottomrule
	\end{tabular}
	}
\vspace{-1.5mm}
\end{table}	

How can we extract a lossless summary from a dynamic graph and update the summary, reflecting changes in forms of edge additions and deletions? 
Can we perform each update in near-constant time while maintaining a concise summary?

Datasets representing relationships between objects are universal in both academia and industry, and graphs are widely used as a simple but powerful representation of such data. For example, graphs naturally represent social networks, WWW, internet topologies, citation networks, and even protein-protein interactions. 
 
With the emergence of big data, real-world graphs have increased dramatically in size, %
and many of them are still evolving over time.
For example, the number of active users in Facebook (i.e., the number of active nodes in the Facebook graph) has increased dramatically from $240$ millions to $2.4$ billions in the last decade. %
Such large dynamic graphs are naturally represented as a {\it fully dynamic graph stream}, i.e., a stream of edge insertions and deletions over time. 
 
To manage large-scale graphs efficiently, compactly representing graphs %
has become important. 
Moreover, compact representations allow a larger portion to be stored in main memory or cache and thus can speed up computation on graphs \cite{buehrer2008scalable, shun2015smaller}. 
Thus, many graph-compression techniques have been proposed: %
relabeling nodes \cite{boldi2004webgraph,chierichetti2009compressing,dhulipala2016compressing}, pattern mining \cite{buehrer2008scalable},
lossless graph summarization \cite{khan2015set,navlakha2008graph,shin2019sweg},
lossy graph summarization \cite{lefevre2010grass,riondato2017graph,beg2018scalable}, to name a few.

Lossless graph summarization is one of the most effective graph compression techniques.
Note that we use the term ``lossless graph summarization'' to indicate this specific compression technique throughout this paper.
Applying lossless graph summarization to a graph $G = (V, E)$ yields  (a) a summary graph $G^* = \summary$, where $S$ is a partition of $V$ (i.e., each element of $S$ is a subset of $V$, and every node in $V$ belongs to exactly one element of $S$) and $P$ is a set of pairs of two elements in $S$, and (b) edge corrections $C=\cor$.
Let $\hat{G}$ be the graph where all pairs of nodes in each pair of subsets in $P$ are connected and the others are not. Then, $C^+$ and $C^-$ are the sets of edges to be added to and removed from $\hat{G}$, respectively, for recovering the original graph $G$.
In addition to yielding compact representations, lossless graph summarization stands out among many compression techniques due to the following desirable properties:

\vspace{-1mm}
\begin{itemize}[leftmargin=*]
	\item \textbf{Queryable}: neighborhood queries (i.e., retrieving the neighborhood of a query node) can be answered rapidly from a summary graph and edge corrections (Lemma~\ref{sec:method:theory:getnbd} in Sect.~\ref{sec:method:theory}). Neighborhood queries are the key building blocks repeatedly called in numerous graph algorithms (DFS, Dijkstra's algorithm, PageRank, etc).
	\item \textbf{Combinable}: summary graphs and edges corrections, which are in the form of graphs, can be further compressed by any graph-compression methods. Combined with \cite{boldi2004webgraph,buehrer2008scalable,chierichetti2009compressing,dhulipala2016compressing}, lossless graph summarization achieves up to $3.4\times$ additional compression \cite{shin2019sweg}. 
\end{itemize}
\vspace{-0.5mm}

The lossless graph summarization problem \cite{navlakha2008graph}, whose goal is to find the most concise representation in the form of a summary graph and edge corrections, %
has been formulated only for static graphs. %
Thus, as shown in Table~\ref{tab:compare}, all existing solutions \cite{khan2015set,navlakha2008graph,shin2019sweg} are prohibitively inefficient for dynamic graphs, represented as fully dynamic graph streams.
Specifically, since these batch algorithms are not designed to allow for incremental changes in the input graph, they should be rerun from scratch to reflect such changes.
	
\begin{figure}[t]
	\vspace{-6.5mm}
	\centering
	\hspace{-3mm}
	\subfigure[Speed]{\label{fig:ad:speed}
		\includegraphics[width=.325\columnwidth]{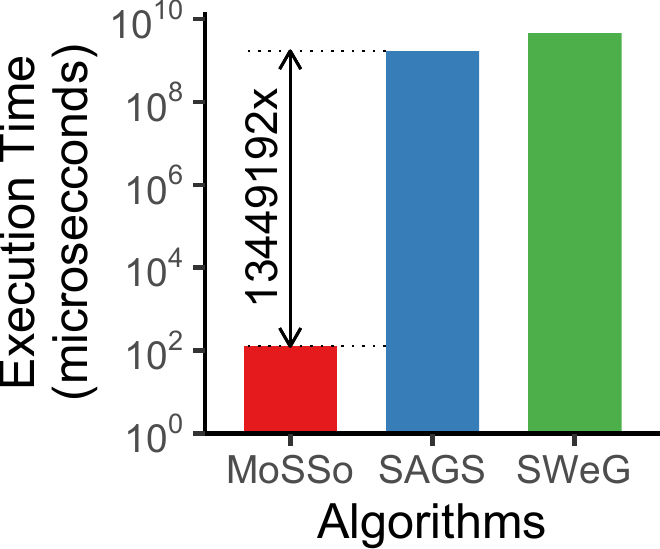}
	} \hspace{-1.5mm}
	\subfigure[Compression]{\label{fig:ad:compress}
		\includegraphics[width=.33\columnwidth]{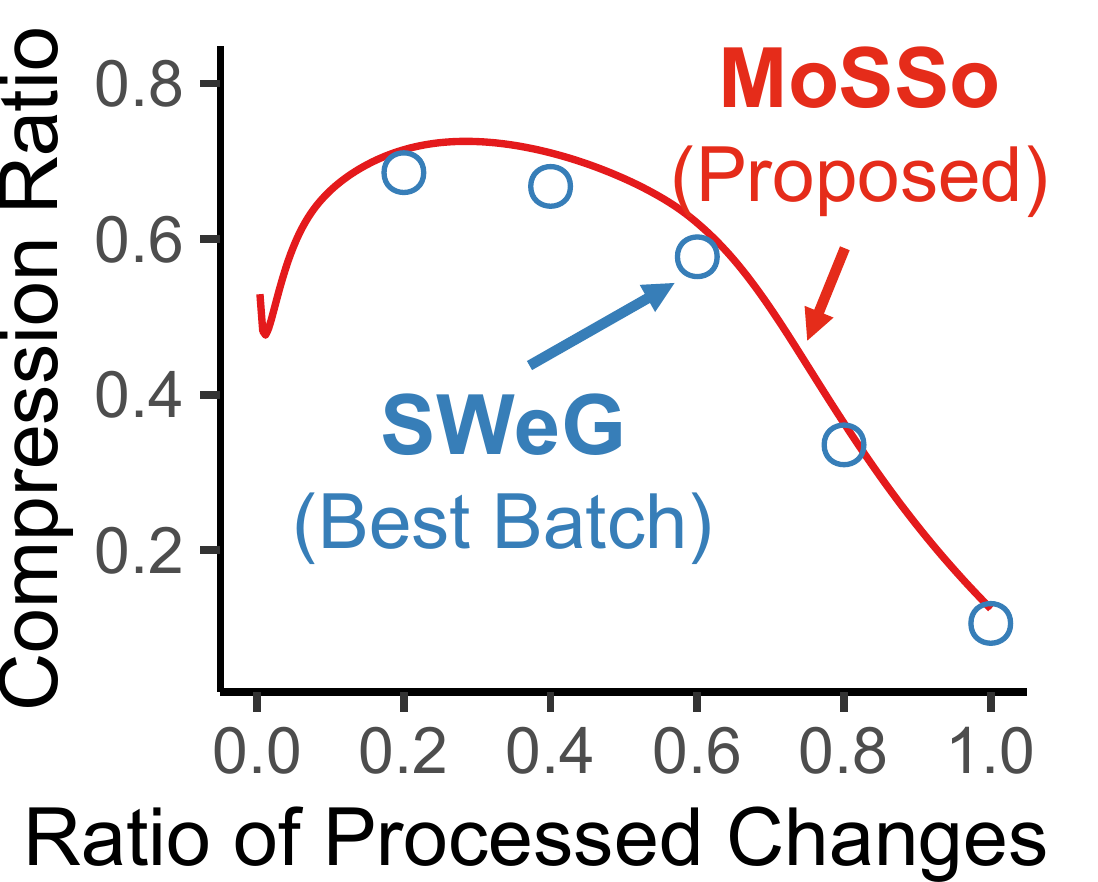}
	} \hspace{-3mm}
	\subfigure[Scalability]{\label{fig:ad:scalable}
		\includegraphics[width=.295\columnwidth]{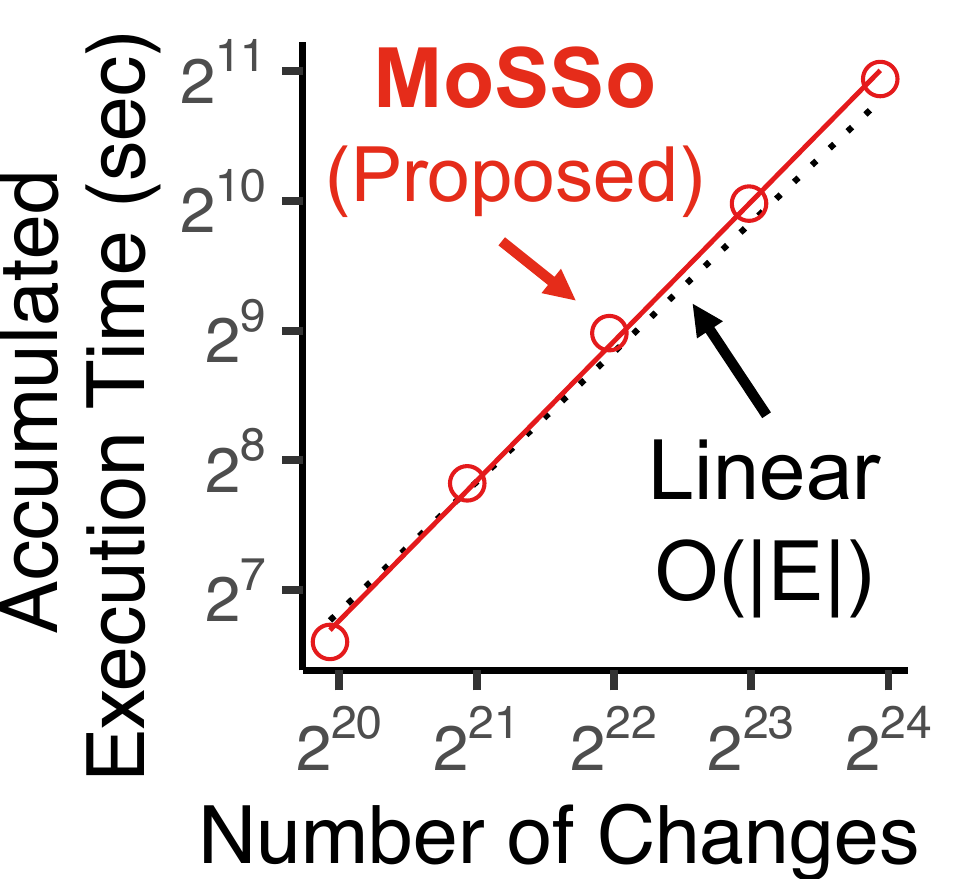}
	} \\
	\vspace{-2.5mm}
	\caption{\label{fig:ad}
		\underline{\smash{\textsc{MoSSo} is fast, effective, and scalable}}.
		(a) Fast: processes a change up to 7 orders of magnitude faster than the fastest batch method.
		(b) Effective: gives comparable compression rates to the best batch method.	
		(c) Scalable: handles each change in near-constant time,  i.e., its total runtime is near-linear in the number of changes. See Sect.~\ref{sec:exp} for details.
	}
\end{figure}

In this work, we formulate a new problem, the goal of which is lossless summarization of a fully dynamic graph. 
Then, we propose the first incremental algorithm \algo (\textbf{Mo}ve if \textbf{S}aved, \textbf{S}tay \textbf{o}therwise)
for the problem. 
In response to each change in the input graph, 
\algo updates the summary graph and edge corrections by moving nodes among supernodes.
(a) \alone: injects randomness in order to escape a local optimum and to cope with changing optima, 
(b) \fastr: performs neighborhood sampling, which is repeatedly executed in \algo, in near-constant time on the output representation without having to retrieve all neighbors,\footnote{Naively, we retrieve all neighbors of a query node from the output representation, which takes time proportional to its degree, and then uniformly sample some.} 
and (c) \guide: when selecting a new supernode into which a node moves, selects it effectively using coarse clustering. Note that the coarse clustering is distinct from graph summarization, which we refer to as fine clustering.

We evaluate \algo with respect to speed and compression rate on $10$ real-world graphs with up to $0.3$ billion edges.
Specifically, we compare \algo with state-of-the-art batch algorithms for lossless graph summarization as well as streaming baselines based on greedy and Markov chain Monte Carlo methods. Through theoretical and empirical analyses, we show the following merits of \algo:

\vspace{-0.5mm}
\begin{itemize}[leftmargin=*]
	\item \textbf{Fast and `any time':} It takes near-constant time to process each change (0.1 millisecond per change), and is up to 10 million times faster than state-of-the-art batch algorithms (Fig.s~\ref{fig:ad:speed}, \ref{fig:ad:scalable}).
	\item \textbf{Scalable:} It requires sub-linear memory (Thm.~\ref{sec:method:theory:thm_space}) while summarizing graph streams with up to hundreds of millions of edges.
	\item \textbf{Effective:} It does not pale in comparison with state-of-the-art batch algorithms in terms of compression rates (Fig.~\ref{fig:ad:compress}).
\end{itemize}
\vspace{-0.5mm}
{\bf Reproducibility}: The code and datasets used in the paper are available at \url{http://dmlab.kaist.ac.kr/mosso/}.

In Sect.~\ref{sec:problem}, we introduce some notations and concept, and we formally formulate a new problem whose goal is to incrementally summarize graph streams. In Sect.~\ref{sec:method}, we present \algo and analyze it theoretically.
In Sect.~\ref{sec:exp}, we provide experimental results. 
After reviewing previous work in Sect.~\ref{sec:related}, we offer conclusions in Sect.~\ref{sec:conclusions}.

\vspace{-1mm}
\section{Notations and Problem Setup}
\vspace{-1mm}
\label{sec:problem}
\begin{table}[t]
	\vspace{-5mm}
	\caption{\label{tab:symbol} Table of Symbols.}
	\footnotesize
	\hspace{-2mm}
	\scalebox{1}{
		\begin{tabularx}{\linewidth} {l|l}
			\toprule
			{\bf Symbol}    &  {\bf Meaning}  \\ 
			\midrule
			\multicolumn{2}{l}{ Symbols for the problem definition (Sect.~\ref{sec:problem:notations})} \\
			\midrule
			$G$=$(V,E)$  & a simple undirected graph with nodes $V$ and edges $E$ \\
			$N(u)$    & neighborhood of node $u$ \\
			$\textup{deg}(u)$ & degree of node $u$  \\
			$\stream$  & a fully dynamic graph stream \\
			$e_t$=$\ea$   & edge insertion at time $t$ \\
			$e_t$=$\ed$   & edge deletion at time $t$ \\
			$G_t$ & $G$ at time $t$ \\
			$G^*$=$\summary$ & a summary graph with supernodes $S$ and superedges $P$\\
			$\Cp$	& set of edges to be inserted \\
			$\Cm$	& set of edges to be deleted \\
			$\varphi$  & objective function that we aim to minimize\\
			\midrule
			\multicolumn{2}{l}{Symbols for the algorithm descriptions (Sect.~\ref{sec:method:term})} \\
			\midrule
			$\target$ & a testing node \\
			$\tpsymb$ & testing pool related to node $u$ \\
			$\tnsymb$ & testing nodes related to node $u$ \\
			$\candidate$ & a candidate \\
			$\cpsymb$ & candidate pool for testing node $\target$ \\
			$E_{AB}$ & edges between supernodes $A$ and $B$ \\
			$T_{AB}$ & all possible edges between supernodes $A$ and $B$ \\
			$S_v$ & supernode in $G^*$ that contains node $v$ \\
			$N(S_v)$ & neighborhood of supernode $S_v$ in a summary graph $G^*$ \\
			$\Cp(u)$ & edges in $\Cp$ incident to node $u$ \\
			$\Cm(u)$ & edges in $\Cm$ incident to node $u$ \\
			\bottomrule       
		\end{tabularx}
	}
\end{table}

\begin{figure}[t]
	\centering
	\vspace{-4mm}
	\includegraphics[width=\linewidth]{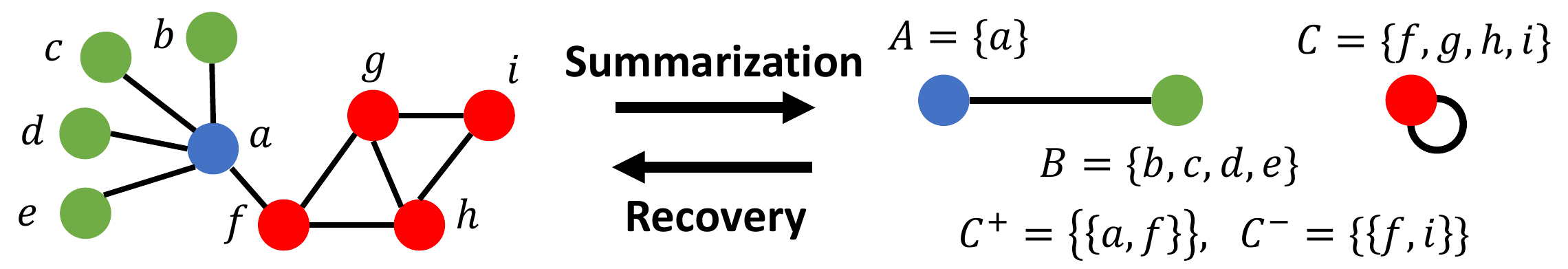}
	\\ \vspace{-1mm}
	\caption{\label{fig:summary} \underline{\smash{Example of lossless graph summarization.}} The input graph has $10$ edges, while a summary graph and edge corrections have only $4$ (super) edges, i.e., $|P|+|\Cp|+|\Cm|=4$. The optimal encoding in Sect.~\ref{sec:method:term} creates the superedge $\{A, B\}$ since $|E_{AB}|>\frac{|T_{AB}|+1}{2}$ but not $\{A, C\}$ since $|E_{AC}|\leq \frac{|T_{AC}|+1}{2}$.}
\end{figure}

We first introduce some important notations and concepts, which are listed and illustrated in Table~\ref{tab:symbol} and Fig.~\ref{fig:summary}.
Then, we formulate a new problem, namely incremental lossless graph summarization.

\vspace{-1mm}
\subsection{Notations and Concepts}
\label{sec:problem:notations}
\vspace{-1mm}

\smallsection{Graph:}
A \textit{graph} $G = (V, E)$ consists of a set $V$ of \textit{nodes} and a set $E$ of \textit{edges}.
We especially consider a simple undirected graph;
each edge $\edge\in E$ is an unordered pair of distinct nodes $u, v \in V$. 
We denote the {\it neighborhood} of a node $u$ (i.e., the set of nodes adjacent to $u$) as $N(u)\subset V$, and we define the degree of $u$ as $|N(u)|$.

\begin{figure*}[t]
	\centering
	\vspace{-5mm}
	\includegraphics[width=\linewidth]{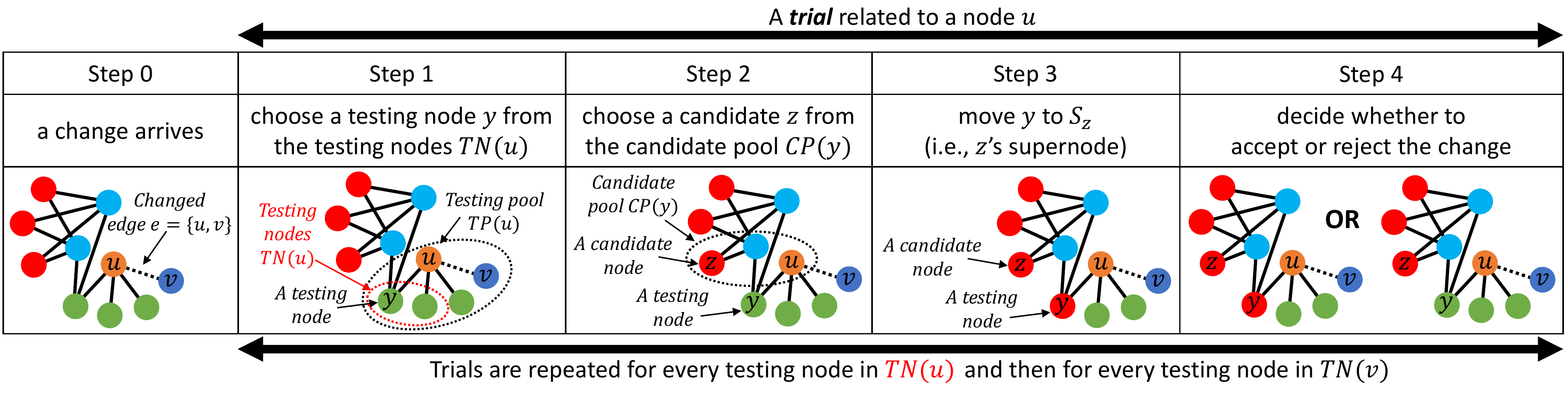}
	\\ \vspace{-2.5mm}
	\caption{\label{fig:trial} \underline{\smash{Example trial to update the summary graph in response to a change $\ea$ or $\ed$.}}
		The color of nodes indicates their membership to supernodes.
		For an input node $u$ (and then for $v$),
		the algorithms presented in Sect.~\ref{sec:method} set a testing pool $\tpsymb$ and obtain testing nodes $\tnsymb\subseteq\tpsymb$.
		Then, they repeat trials (i.e., the above steps 1-4) for every testing node in $\tnsymb$.
		Note that the algorithms set $\tpsymb$, $\tnsymb$, and $\cpsymb$ in different ways.}
\end{figure*}

\smallsection{Summary Graph:}
A graph $G^*=\summary$ is a \textit{summary graph} of a graph $G = (V, E)$ if its node set $S$ is a partition of $V$, %
i.e., if $V = \bigcup_{\alpha} S_{\alpha}$ with $S_{\alpha} \cap S_{\beta} = \emptyset$ for any distinct $S_{\alpha}$, $S_{\beta} \in S$. 
To avoid confusion, we call nodes and edges in $G$ simply \textit{nodes} and  \textit{edges}, while we call them in its summary graph $G^*$ \textit{supernodes} and \textit{superedges}. 
We denote the supernode containing a node $u$ as $S_u$ and the superedge between two supernodes $S_u, S_v\in S$ as $\{S_u, S_v\}\in P$.

\smallsection{Edge Corrections \& Recovery:}
Given a summary graph $G^*=\summary$ of a graph $G=(V,E)$, we obtain a graph $\hat{G}=(V,\hat{E})$ by connecting every pair of nodes between two neighboring supernodes, i.e., $\{u,v\}\in \hat{E}$ if and only if $u\neq v$ and $\{S_u,S_v\}\in P$.
We can say $G^*$ roughly describes $G$ if $\hat{G}$ is similar to $G$.
Moreover, with \textit{edge corrections} $C = \cor$ where $\Cp:=E-\hat{E}$ and $\Cm:=\hat{E}-E$, the original graph $G=(V,E)$ is exactly recovered from $G^*$ as follows:
\begin{equation*}
	V \leftarrow \bigcup\nolimits_{S_\alpha \in S}S_{\alpha}, \ \ \ \ E \leftarrow (\hat{E} \cup \Cp) \setminus \Cm.
\end{equation*}
That is, $G^*$ and $C$  losslessly summarize $G$.
Given $G$, the {\it lossless summarization problem} \cite{navlakha2008graph} is to find the most concise $G^*$ and $C$.

\smallsection{Fully Dynamic Graph Stream:}
We define a \textit{fully dynamic graph stream} by %
a sequence $\stream$ of changes. 
Each change $e_t$ is either an \textit{edge addition} $e_t=\ea$ or an \textit{edge deletion} $e_t=\ed$ between distinct nodes $u,v\in V$.
For each edge addition $e_t=\ea$, $u$ and/or $v$ may be new nodes unseen until the current time $t$.
Since node additions and deletions are expressed as a series of the additions and deletions of their adjacent edges, respectively,
a sequence of edge additions and deletions is %
expressive enough to represent a dynamic graph with new and deleted nodes.

An empty graph $G_0=\emptyset$ evolves according to $\stream$. We define the graph at time $t$ by $G_t = (V_t, E_t)$, where $V_t$ and $E_t$ are obtained in the inductive way: 
(1) {\bf addition:} if $e_{t-1}=\ea$, then $E_t = E_{t-1} \cup \{\edge\}$ and $V_t = V_{t-1} \cup \edge$, (2) {\bf deletion:} if $e_{t-1}=\ed$, then $E_t = E_{t-1} \setminus \{\edge\}$ and $V_{t}= V_{t-1}$. %
We assume $\edge \notin E_{t-1}$ for every edge addition and $\edge \in E_{t-1}$ for every edge deletion. That is, we assume that the graph stream is sound.

\vspace{-1mm}
\subsection{Problem Formulation}
\label{sec:problem:definition}
\vspace{-1mm}

Given a fully dynamic graph $G_t$ evolving under $\stream$, the most concise summary graph $\GISTAR$ and edge corrections $\CI$ change in response to each change $e_{t}$. Since recomputing concise $\GISTAR$ and $\CI$ for each $G_t$ from scratch is prohibitive for large-scale graphs, an incremental update on $\GISTAR$ and $\CI$ to obtain concise $\GIPSTAR$ and $\CIP$ is highly desirable. Thus, we formulate a new problem, namely {\it incremental lossless graph summarization}, in Problem~\ref{problem}.

\noindent\fbox{%
	\parbox{0.97\columnwidth}{%
		\vspace{-1.5mm}
		\begin{problem}\normalfont{(Incremental Lossless Graph Summarization).}\label{problem}
			\vspace{-1mm}
			\begin{enumerate}[leftmargin=*]
				\item \textbf{Given:} a fully dynamic graph stream $\stream$ %
				\item \textbf{Retain:} a summary graph $\GISTAR = \summaryt$ and edge corrections $C=\cort$ of graph $G_t$  at current time $t$
				\item \textbf{To Minimize:} the size of output representation, i.e.,
				\vspace{-1mm}
				\begin{equation}
				\varphi(t) := |P_t|+|C_t^+|+|C_t^-| \label{eq:objective}
				\end{equation}
				\vspace{-2mm}
			\end{enumerate}
			\vspace{-2mm}
		\end{problem}
		\vspace{-3mm}
	}%
}
\vspace{0.1mm}

The objective $\varphi(t)$ measures the size of the summary graph $\GISTAR$ and the edge corrections $\CI$. In $\varphi(t)$, the number of supernodes, which is marginal compared to that of (super) edges appearing in output representations, is disregarded for simplicity. This objective was first proposed by \cite{navlakha2008graph} under the \textit{Minimum Description Length} principle \cite{rissanen1978modeling}, which originates from information theory. %
Then, a number of algorithms for lossless summarization of static graphs have been developed based on the objective \cite{navlakha2008graph,khan2015set,shin2019sweg}.

In Problem~\ref{problem}, there are almost infinitely many possible summary graphs   \cite{bell1938iterated} and moreover, the optimal solution is subject to change at any time.
Thus, we focus on streaming algorithms for finding approximate solutions. 
Note that, even for static graphs, only heuristics without guarantees have been studied.

\vspace{-1mm}
\section{Proposed Method}
\vspace{-1mm}
\label{sec:method}

In this section, we propose \algo, a fast and space-efficient streaming algorithm for lossless graph summarization.
We also present baseline streaming algorithms and a preliminary version of \algo, while overcoming their drawbacks addressed in full-fledged \algo. 

We first introduce terms and subroutines used in the algorithms. 
Then, we present the baseline algorithms. %
After that, we present a preliminary version of \algo and then full-fledged \algo, with a focus on ideas that leads to improved scalability and compression rates. 
Lastly, we analyze the time and space complexities of \algo.

\vspace{-1mm}
\subsection{Common Terms and Subroutines (Fig.~\ref{fig:trial})}
\label{sec:method:term}
\vspace{-1mm}

We define some terms and subroutines commonly used in the algorithms presented in the following subsections.

\smallsection{Common Terms:}
Assume that $\edge$ is inserted or deleted at current time $t$ (i.e., $e_t=\ea$ or $\ed$).
We call nodes $u$ and $v$ \textit{input nodes}.
A \textit{trial related to} $u$ on a node $x$ is the attempt to change the supernode membership $S_x$ of the node $x$, which is somehow related to $u$. 
A \textit{\tp} $\tpsymb$ is the set of nodes on which a trial related to $u$ can be proceeded. %
A \textit{\tn} is a node on which a trial is actually proceeded at current time $t$.
We let $\target$ be each testing node and $\tnsymb\subseteq \tpsymb$ be the set of all testing nodes at current time $t$. 
A \textit{\cand} $\candidate$ is a node in the supernode into which the \tn $\target$ tries moving. Note that this move can be rejected and reverted.
A \textit{\cp} $\cpsymb\subseteq V$ is a set of all possible candidates given to the \tn $\target$.

\begin{figure*}[t]
	\vspace{-4.2mm} %
\end{figure*}

\smallsection{Trial:}
A subroutine called
\textit{trial related to $u$} is described as follows:

\noindent\fbox{%
	\parbox{0.97\columnwidth}{%
		{\bf A trial related to $u$ (the terms defined above are used):}
		\vspace{-0.5mm}
		\begin{enumerate}[leftmargin=*]
			\item Select a \textit{\tn} $\target$ from $\tnsymb$.
			\item Select a \textit{\cand} $\candidate$ from $\cpsymb$.
			\item Propose moving $y$ into $\candidateS$.
			\item Determine whether to accept or reject the move, based on the change in the objective $\varphi$ (when assuming the optimal encoding described in the next paragraph).
		\end{enumerate}
		\vspace{-1mm}
	}%
}
\vspace{0.2mm}

\noindent This reflects why we name the final algorithm \algo (\textbf{Mo}ve if \textbf{S}aved, \textbf{S}tay \textbf{o}therwise). 
This trial is repeated for every \tn in $\tnsymb$.
The algorithms presented in the following subsections are distinguished by how they (1) set a \tp $\tpsymb$, (2) extract testing nodes $\tnsymb$ from $\tpsymb$, (3) set a \cp $\cpsymb$, (4) propose a \cand from $\cpsymb$, and (5) accept the proposal.

\smallsection{Optimal Encoding:}
While finding the \textit{optimal} set $S$ of supernodes, which minimizes the objective $\varphi$, is challenging,
finding optimal $P$ and $C$ for current $S$ is straightforward. For each supernode pair $\{A,B\}$, let $E_{AB} := \{ \edge \in E | u\in A, v \in B, u\neq v\}$ and $T_{AB} := \{ \edge \subseteq V  | u\in A, v \in B, u\neq v\}$ be the sets of existing and potential edges between $A$ and $B$, respectively.
Then, the edges between $A$ and $B$ (i.e., $E_{AB}$) are optimally encoded as follows:

\noindent\fbox{%
	\parbox{0.97\columnwidth}{%
		{\bf Optimal encoding for the edges in $E_{AB}$:}
		\vspace{-0.5mm}
		\begin{enumerate}[leftmargin=*]
			\item If $|E_{AB}| \leq \frac{|T_{AB}| + 1}{2}$, then add all edges in $E_{AB}$ to $\Cp$.
			\item If $|E_{AB}|>\frac{|T_{AB}| + 1}{2}$, then add the superedge $\{A, B\}$ to $P$ and $T_{AB} \setminus E_{AB}$ to $\Cm$.
		\end{enumerate}
	\vspace{-1mm}
	}%
}

\noindent Note that adding all edges in $E_{AB}$ to $\Cp$ increases $\varphi$ by $|E_{AB}|$, while adding the superedge $\{A, B\}$ to $P$ and $T_{AB} \setminus E_{AB}$ to $\Cm$ increases $\varphi$ by $1 + |T_{AB}| - |E_{AB}|$.
The above rules always choose an option resulting in a smaller increase in $\varphi$.

\vspace{-1mm}
\subsection{\sgreedy: First Baseline}
\label{sec:method:sgreedy}	
\vspace{-1mm}

We present \sgreedy, a baseline streaming algorithm for Problem~\ref{problem}, and then we point out its limitations.

\smallsection{Procedure:}
When an edge $\edge$ is inserted or deleted, \sgreedy greedily moves $u$ and then $v$, while fixing the other nodes, so that the objective $\varphi$ is minimized.
That is, in terms of the introduced notions, %
\sgreedy sets $\tpsymb=\tnsymb=\{u\}$, and a candidate is chosen from $\cpsymb=V$ so that  $\varphi$ is minimized.

\smallsection{Limitation:}
While $|TN(u)|$ is just $1$, unlike the other algorithms described below, this approach is computationally expensive as it takes all supernodes into account to find a locally best \cand. It is also likely to get stuck in a local optimum, as described below.

\vspace{-1mm}
\begin{itemize}[wide]
	\item \textit{Limitation 1} (\fprob): \sgreedy lacks exploration for reorganizing supernodes, and thus nodes tend to stay in supernodes that they move into in an early stage. This stagnation also prevents new nodes from moving into existing supernodes. These lead to poor compression rates in the long run.
\end{itemize}
\vspace{-1mm}

\vspace{-1mm}
\subsection{\mcmc: Second Baseline}
\label{sec:method:mcmc}
\vspace{-1mm}

We present \mcmc, another streaming baseline algorithm based on randomized search.
It significantly reduces the computational cost of each trial, compared to \sgreedy, since it does not have to find the best candidate, and thus makes more trials affordable. 
Moreover, its randomness helps escaping from local optima and smoothly coping with changing optima.
However, this approach also suffers from two drawbacks, as described later.

\smallsection{Motivation:}
Randomized searches based on Markov Chain Monte Carlo (MCMC) \cite{hastings1970monte} have proved effective for the inference of stochastic block models (SBM) \cite{kao2017streaming, peixoto2014efficient}.
We focus on an interesting relation between communities in SBM and supernodes in graph summarization. 
Since nodes belonging to the same community are likely to have similar connectivity, 
grouping them into a supernode may achieve significant reduction in $\varphi$.
Hence, inspired  by \cite{kao2017streaming}, we propose \mcmc for Problem~\ref{problem}. 

\smallsection{Procedure:}
In response to each change $\ea$ \textup{or} $\ed$ in the input graph,
\mcmc performs the following steps for $u$ (and then the exactly same steps for $v$):

\noindent\fbox{%
	\parbox{0.97\columnwidth}{%
		{\bf Trials related to $u$ by \mcmc:}
		\vspace{-0.5mm}
		\begin{enumerate}[leftmargin=*]
			\item Set $\tnsymb = \tpsymb = N(u)$.
			\item For each $\target$ in $\tnsymb$, select a \cand $z$ from $\cpsymb = V$ through sampling according to a predefined proposal probability distribution \cite{peixoto2014efficient}.
			\item For each $\target$, accept the proposal (i.e., move $y$ into $S_{z}$) with an acceptance probability, which depends on the change in $\varphi$. 
		\end{enumerate}
		\vspace{-1mm}
	}%
}

\noindent 
In Step (1), the neighbors of $u$ are used as testing nodes since they are affected most by the input change. The $\textup{deg}(u)$ trials can be afforded since a trial in \mcmc is computationally cheaper than that in \sgreedy.
The proposal probability distribution and the acceptance probability used in Steps (2) and (3) are described in detail, with a pseudo code of \mcmc, in Appendix~\ref{appendix:mcmc}.

\smallsection{Limitations:}
\mcmc suffers from two limitations, which are the bottlenecks of its speed and compression rates.

\vspace{-1mm}
\begin{itemize}[wide]
	\item \textit{Limitation 2} (\sprob): To process each change $\ea$ \textup{or} $\ed$ in the input stream, \mcmc retrieves the neighborhood of many nodes from  current $G^{*}$ and $C$ (see the proof of Lemma~\ref{sec:method:theory:getnbd} for a detailed procedure of the retrieval). 
	Specifically, it retrieves the neighborhood of $u$ in Step (1), and for each testing node $y$, it retrieves the neighborhood of at least one node to select a candidate in Step (2).
	That is, at least $2+|TN(u)|+|TN(v)| = 2 + \textup{deg}(u) + \textup{deg}(v)$ neighborhood retrievals occur.
	Thus, the time complexity of \mcmc is deadly affected by growth of graphs, which may lead to the appearance of high-degree nodes and the increase of average degree \cite{leskovec2007graph}.
	\item \textit{Limitation 3} (\tprob): For proposals to be accepted, promising candidates leading to reduction in $\varphi$ need to be sampled from the proposal probability distribution.
	However, the probability distribution \cite{peixoto2014efficient}, which proved successful for SBM, results in mostly rejected proposals and thus a waste of computational time.
\end{itemize}
\vspace{-1mm}

\vspace{-1mm}
\subsection{\algobasic: Simple Proposed Method}
\label{sec:method:algobasic}
\vspace{-1mm}

We present \algobasic, a preliminary version of \algo, with three novel ideas for addressing the limitations that the baseline streaming algorithms suffer from. 
Then, in order to look for further improvement, we identify some limitations of \algobasic.

\smallsection{Procedure:}
\algobasic is described in Alg.~\ref{algo:algo}.
In response to each change $\ea$ \textup{or} $\ed$ in the input graph,
\algobasic, equipped with the novel ideas described below, conducts the following steps for $u$ (and then exactly the same steps for $v$):

\noindent\fbox{%
	\parbox{0.97\columnwidth}{%
		{\bf Trials related to $u$ by \algobasic:}
		\vspace{-0.5mm}
		\begin{enumerate}[leftmargin=*]
			\item Sample a fixed number (denoted by $c$) of nodes from $N(u)$ and use them as $\tpsymb$.
			\item Add each $w\in\tpsymb$ to $\tnsymb$ with probability $\frac{1}{\textup{deg}(w)}$.
			\item For each $\target\in TN(u)$, with probability $\escape$ of escape, propose creating a singleton supernode $\{\target\}$.
			\item Otherwise, randomly select a \cand $\candidate$ from $\cpsymb$ where $\cpsymb = N(u)$ for every $\target \in \tnsymb$.
			\item For each $\target$, accept the proposal (i.e., move $y$ to $\candidateS$) if and only if it reduces $\varphi$.
		\end{enumerate}
	\vspace{-1mm}
	}%
}

\begin{figure*}[t]
	\vspace{-4.2mm} %
\end{figure*}

\smallsection{Key Ideas:}
Contrary to \mcmc, for each change $\ea$ or $\ed$, \algobasic (1) extracts $\tnsymb$ from $\tpsymb$ probabilistically depending the degrees of nodes, (2) limits $\cpsymb$ to $N(u)$ for every \tn $\target \in \tpsymb$, and (3) occasionally separates $y$ from $S_{y}$ and create a singleton supernode $\{y\}$. 
As shown in Sect.~\ref{sec:exp}, these ideas enable \algobasic to significantly outperform \mcmc, as well as \sgreedy, in terms of speed and compression rates.
Below, we describe each idea in detail.

\vspace{-1mm}
\begin{itemize}[wide]
	\item \textbf{\guide (1)}: When forming $\tnsymb$,
	\algobasic first samples with replacement a fixed number of nodes from $N(u)$ and construct $\tpsymb$ using them. Then, it adds each sampled node $w\in \tpsymb$ to $\tnsymb$ with probability $1/\textup{deg}(w)$.
	In practice, high-degree nodes tend to have unique connectivity, and thus they tend to form singleton supernodes.
	Therefore, moving them rarely leads to the reduction of $\varphi$.
	However, high-degree nodes are frequently contained in $TP(u)$ since %
	edge changes adjacent to any neighbor $u$ put the high-degree nodes into $TP(u)$. %
	Moreover, once they are chosen as testing nodes, computing the change in $\varphi$ and updating the optimal encoding are computationally expensive since they have many neighbors.
	By probabilistically filtering out high-degree nodes when forming $\tnsymb$, \algobasic significantly reduces redundant and computationally expensive trials and thus partially addresses Limitation~3 (i.e., \tprob).

	\item \textbf{\alone}: Instead of always finding a \cand from $\cpsymb$, it separates $\target$ from $S_{y}$ and creates a singleton supernode $\{\target\}$ with probability $\escape \in [0, 1)$. 
	By injecting flexibility to the formation of a summary graph, %
	this idea, which we call \alone, helps supernodes to be reorganized in different and potentially better ways in the long run. 
	Therefore, this idea addresses Limitation~1 and yields significant improvement in compression.

	\item \textbf{\fastr (1)}: By limiting $\cpsymb$ to $N(u)$ for every \tn $\target \in \tnsymb$, \algobasic reduces the number of required neighborhood retrievals and thus partially addresses Limitation~2 (i.e., \sprob). 
	For each input change,
	while \mcmc repeats neighborhood retrievals $2 + \textup{deg}(u) + \textup{deg}(v)$ times,
	\algobasic only retrieves $N(u)$ and $N(v)$.
	Moreover, since $N(u)$ still contains promising candidates, limiting $\cpsymb$ to $N(u)$ does not impair the compression rates as shown empirically in Sect.~\ref{sec:exp}.
\end{itemize}
\vspace{-1mm}

\begin{algorithm}[t]
	\small
	\caption{\label{algo:algo}{\sc \textcolor{red}{\algo} and \textcolor{blue}{\algobasic}}: proposed algorithms for Problem~\ref{problem}.	
	}
	\DontPrintSemicolon
	\nonl $\triangleright$ The \textcolor{red}{red} and \textcolor{blue}{blue} lines are executed only in \textcolor{red}{\algo} and \textcolor{blue}{\algobasic}, respectively.\;
	\nonl $\triangleright$ The black lines are executed in both algorithms. \;
	\KwIn{summary graph: $\GISTAR$, edge corrections: $\CI$, \\ \quad\quad\quad input change: $\{src, dst\}^{+}$ or $\{src, dst\}^{-}$\\ \quad\quad\quad escape probability: $e$, sample number: $c$,
		\textcolor{red}{coarse clusters: $R$}}
	\KwOut{summary graph: $\GIPSTAR$, edge corrections: $\CIP$}
	\textcolor{red}{update $R$ in response to the input change}\;
	\ForEach{$u$ \textup{in} $\textup{\{}src, dst\textup{\}}$} {
		\textcolor{red}{$\tpsymb \gets \getrnd(c)$} \hfill	$\triangleright$ see Alg.~\ref{algo:neighbor}\;
		\nonl \textcolor{blue}{$\tpsymb \gets$ randomly chosen $c$ neighbors from $N(u)$}\;
		\ForEach{$w$ \textup{in} $\tpsymb$}{
			put $w$ into $\tnsymb$ with probability $1/\textup{deg}(w)$ \; 
		}
		\ForEach{$\target$ \textup{in} $\tnsymb$}{
			sample $X \sim uniform(0, 1)$\;
			\eIf{$X \leq e$} {
				temporarily create a new supernode $S_y = \{\target\}$\;
			} {
				\textcolor{red}{$R(\target) \gets$ the coarse cluster containing $\target$}\;
				\textcolor{red}{$\cpsymb \gets \tpsymb \cap R(\target)$} or
				\textcolor{blue}{$\cpsymb \gets N(u)$}\;
				randomly choose a node $\candidate$ from $\cpsymb$\;
				temporarily move $\target$ into $\candidateS$\;
			}	
			compute $\Delta\varphi$ (i.e., change in $\varphi$)  for the proposal\;
			\If{$\Delta\varphi \leq 0$} {
				accept the proposal and update $\GISTAR$, $\CI$\;
			\nonl	\hfill $\triangleright$ see Sect.~\ref{sec:method:term} for optimal encoding
			}
		}
	}
	$\GIPSTAR, \CIP \gets \GISTAR, \CI$\;
	\Return{$\GIPSTAR$, $\CIP$}\;
\end{algorithm}

\begin{figure*}[t]
	\vspace{-4.2mm} %
\end{figure*}

\smallsection{Limitations:}
Although the above ideas successfully mitigate the limitations, there remain issues to be resolved.

\vspace{-1mm}
\begin{itemize}[wide]
	\item \textit{Limitation 2} (\sprob): While \algobasic reduces the number of neighborhood retrievals to two per each input change, the retrievals still remain as a scalability bottleneck. %
	As analyzed in Lemma~\ref{sec:method:theory:getnbd}, retrieving the neighborhood of a node from current $G^{*}$ and $C$ takes $O(\overline{deg})$ time on average\footnote{To retrieve the neighborhood  $N(\target)$ of $\target$, we need to collect all its neighbors in $\Cp$ (i.e., $\{v \in V | \{v, \target\} \in \Cp\}$) and all nodes in the neighboring supernodes of $S_{\target}$ (i.e. $\{ v\in V | v \in S_u,  S_u \in N(S_{\target}) \}$ where $N(S_{\target}) := \{S_u \in S | \{S_u, S_{\target}\} \in P\}$).
	Then, we need to filter out all its neighbors in $\{v \in V | \{u, \target\} \in \Cm\}$ (see Sect.~\ref{sec:problem:notations}).}, 
	where $\overline{deg} = \frac{2|E|}{|V|}$ is the average degree in the input graph.
	However, it is well known that lots of real-world graphs are densified over time \cite{leskovec2007graph}.
	Specifically, the number of edges increases super-linearly in the number of nodes, leading to the growth of $\overline{deg}$ over time. 
	Hence, full neighborhood retrievals pose a huge threat to scalability.
	
	\item \textit{Limitation 3} (\tprob): While \algobasic uses the degree of nodes to reduce redundant trials, which lead to rejected proposals, it does not fully make use of structural information around input nodes but simply draws a random \cand from $N(u)$. 
	Careful selection of candidates based on the structural information 
	can be desirable to further reduce the number of redundant trials and thus to achieve concise summarization rapidly. 

\end{itemize}

\vspace{-1mm}
\subsection{\algo: Full-Fledged Proposed Method}
\label{sec:method:algo}
\vspace{-1mm}

We present \algo, the full-fledged version of our algorithms.
To overcome the aforementioned drawbacks of \algobasic, \algo employs (1) coarse clustering for careful candidate selection and (2) \getrnd, a novel sampling method instead of full neighborhood retrievals. 
Equipped with these ideas, \algo achieves near-constant processing time per change and compression rates even comparable to state-of-the-art batch algorithms.

\smallsection{Procedure:}
A pseudo code of \algo is provided in Alg.~\ref{algo:algo}.
In response to each change $\ea$ \textup{or} $\ed$,
\algo conducts the following steps for $u$ (and then exactly the same steps for $v$):

\noindent\fbox{%
	\parbox{0.97\columnwidth}{%
		{\bf Trials related to $u$ by \algo:}
		\vspace{-0.5mm}
		\begin{enumerate}[leftmargin=*]
			\item Update \textbf{coarse clusters} in response to the change.
			\item Sample a fixed number (denoted by $c$) of nodes from $N(u)$, \textbf{without retrieving all} $N(u)$, and use them as  $\tpsymb$.
			\item Add each $w\in\tpsymb$ to $\tnsymb$ with probability $\frac{1}{\textup{deg}(w)}$.
			\item For each $\target\in TN(u)$, with probability $\escape$ of escape, propose creating a singleton supernode $\{\target\}$.
			\item Otherwise, randomly select a \cand $\candidate$ from $\cpsymb$ where $\cpsymb = \tpsymb \cap R(\target)$ and $R(\target)$ is the \textbf{coarse cluster} containing $\target$, for every $\target \in \tnsymb$.
			\item For each $\target$, accept the proposal (i.e., move $y$ to $\candidateS$) if and only if it reduces $\varphi$.
		\end{enumerate}
	\vspace{-1mm}
	}%
}

\smallsection{Key Ideas:}
The gist of \algo consists of two parts: (1) rapidly and uniformly sampling neighbors from $N(u)$ without retrieving the entire $N(u)$ from $G^{*}$ and $C$ and (2) using an online coarse clustering to narrow down $\cpsymb$.
Below, we describe each idea in detail.

\vspace{-1mm}
\begin{itemize}[wide]
	\item \textbf{\fastr (2)}: As explained in Limitation 2 (Sect.~\ref{sec:method:algobasic}), for scalability, it is inevitable to devise a neighborhood sampling method less affected by the average degree, which tends to increase over time. 
	Thus, we come up with \getrnd, described in Alg.~\ref{algo:neighbor}.
	It is an MCMC method for rapidly sampling nodes from $N(u)$ in an unbiased manner without retrieving the entire $N(u)$. 
	After obtaining a sufficient number of neighbors by \getrnd, \algo limits $\tpsymb$ to the sampled neighbors. %

	Assume that the neighborhood in $\Cp$, $\Cm$ and $P$ of each node is stored in a hash table, and
	let $N(S_{u}) := \{S_v \in S | \{S_u, S_{v}\} \in P\}$ be the set of neighboring supernodes of a supernode $S_{u}$.
	Then, $v\in N(u)$ can be checked rapidly as follows:
	
	\noindent\fbox{%
		\parbox{0.97\columnwidth}{%
			{\bf Checking $v\in N(u)$ on $G^{*}$ and $C$:}
			\vspace{-0.5mm}
			\begin{enumerate}[leftmargin=*]
				\item If $v\in C^{-}(u)$, then $v\notin N(u)$.
				\item Else if $v\in C^{+}(u)$ or $S_{v}\in N(S_{u})$, then $v\in N(u)$.
				\item Else $v\notin N(u)$.
			\end{enumerate}
		}%
	\vspace{-1mm}
	}
	
	\noindent The neighbors of $u$ are divided into two disjoint sets: (1) those in $\Cp(u)$ and (2) those in any supernode in $N(S_u)$. 
	Then, we can uniformly sample a neighbor of $u$ by uniformly sampling a node either from the first set (with probability $\frac{|\Cp(u)|}{\textup{deg}(u)}$) or from the second set (with probability $1-\frac{|\Cp(u)|}{\textup{deg}(u)}$).
	Since the first set is already materialized (by our assumption), uniform sampling from  it is straightforward.
	The remaining challenge is to uniformly sample a node from the second set without materializing the second set. 
	This challenge is formulated in Problem~\ref{problem:neighbor}, where we denote $N(S_u)$ by $\{S_1,..., S_k\}$.

	\noindent\fbox{%
		\parbox{\columnwidth}{%
			
			\vspace{-2mm}
			\begin{problem}[Subproblem 1 for \fastr (2)]\label{problem:neighbor} We have $k$ disjoint supernodes $S_1,\cdots, S_k$ with $N_i := N(u) \cap S_i$ for each $i$.
			For large $k$, it is computationally expensive to obtain $\bigcup_{i=1}^k N_i$.
			However, we can rapidly check whether a given node is contained in $\bigcup_{i=1}^k N_i$.
			How can we rapidly and uniformly draw nodes from $\bigcup_{i=1}^k N_i$?
			\end{problem}
		\vspace{-2mm}
		\textbf{Our solution:} 
			Instead of uniform sampling from $\bigcup_{i=1}^k N_i$, \getrnd, described in Alg.~\ref{algo:neighbor}, samples a node from $\bigcup_{i=1}^k S_i$ and retries sampling if the node is not in $\bigcup_{i=1}^k N_i$. To this end, it randomly selects a supernode $S_i$ with probability
			\begin{equation}
			\pi(S_i \textup{ is selected}) := {|S_i|}/({|S_1| + \cdots + |S_k|}),\label{eq:probdist} 
			\end{equation}
			and then draws a random node from the selected supernode. If the node is not in $\bigcup_{i=1}^k N_i$, then this procedure is repeated again from beginning. It is guaranteed that this sampling scheme draws each node in $\bigcup_{i=1}^k N_i$ uniformly with probability $1/N$, where $N := \sum_{i=1}^k |N_i|$, as shown in Thm.~\ref{sec:method:theory:thm1} in Sect.~\ref{sec:method:theory}.
		}%
	}

	In the above solution, when sampling supernodes according to Eq.~\eqref{eq:probdist}, it is desirable to avoid computing $|S_i|$ for every $i\in\{1,...,k\}$, since $k$ can be large, as formulated in Problem~\ref{problem:mcmc}.

	\noindent\fbox{%
		\parbox{\columnwidth}{%
			\vspace{-2mm}
			\begin{problem}[Subproblem 2 for \fastr (2)] \label{problem:mcmc} How to sample supernodes according to Eq.~\eqref{eq:probdist} without computing $|S_i|$ for each $i$?
			\end{problem}
			\vspace{-2mm}
			\textbf{Our solution:} 
			\getrnd employs MCMC, which basically constructs a Markov chain asymptotically equal to Eq~\eqref{eq:probdist}.
			Specifically, a supernode $S_{\textup{p}}$ is proposed uniformly at random among $k$ supernodes, and then it replaces a previously sampled supernode, which is denoted by $S_{\textup{n}}$, with the following acceptance probability:
			$$
			\textup{min}\left(1, \frac{\pi(S_{\textup{p}}\textup{ is selected})}{\pi(S_{\textup{n}} \textup{ is selected})}\right) = \textup{min}\left(1, \frac{|S_{\textup{p}}|}{|S_{\textup{n}}|}\right).$$ 
			The soundness is shown in Thm.~\ref{sec:method:theory:thm2} in Sect.~\ref{sec:method:theory}.
		}%
	}

	\noindent 
	Both solutions are combined in Alg.~\ref{algo:neighbor}, which describes the entire process for sampling $c$ neighbors from $N(u)$. We analyze 
	its soundness and time complexity in Sect.~\ref{sec:method:theory}.

	\item \textbf{\guide (2)}: 
	To choose candidates leading to a significant reduction in $\varphi$, \algo uses coarse clusters, each of which consists of nodes with similar connectivity.
	Specifically, in \algo, the candidate pool $CP(y)$ of each \tn $y$ consists only of nodes belonging to the same cluster of $y$.
	The coarse clusters are distinct from supernodes, which can be thought as fine clusters.
	
	Any incremental graph clustering methods \cite{ning2007incremental,takaffoli2013incremental,zhao2019incremental,zhou2010clustering} can be used for coarse clustering.
	We especially use min-hashing \cite{broder2000min} as it is fast with desirable theoretical properties: the probability that two nodes belong to the same cluster is proportional to the jaccard similarity of their neighborhoods.
	Moreover, clusters grouped by min-hashing can be updated rapidly in response to changes.

\end{itemize}

By employing the above ideas, \algo processes each change in near-constant time and shows outstanding compression, as shown theoretically in the next subsection and empirically in Sect.~\ref{sec:exp}.

\begin{algorithm}[t]
	\small
	\DontPrintSemicolon
	\KwIn{summary graph: $\GISTAR$, edge corrections: $\CI$, \\ \quad\quad\quad input node: $u$, sample number: $c$}
	\KwOut{list of $c$ random neighbors from $N(u)$}
	$V_r \gets [$ $]$, $P \gets N(S_u), Cp \gets \Cp_t(u), Cm \gets \Cm_t(u)$\;
	$S_n \gets$ random supernode in $P$\; 
	\While {$|V_r| < c$} {
		sample $X_1 \sim uniform(0, 1)$\;
		\eIf{$X_1 \leq |Cp| / \textup{deg}(u)$} {
			put a random node from $Cp$ into $V_r$\;
		} {
			\While {} {
				$S_p \gets$ a random supernode in $P$\;
				sample $X_2 \sim uniform(0, 1)$\;
				\If {$X_2 \leq \textup{min}(1, |S_p|/|S_n|)$} {
					$S_n \gets S_p$\;
				}
				$w \gets$ a random node in $S_n$\;
				\If {$w \notin Cm$} {
					add $w$ to $V_r$\;
					\textbf{break}\;
				}
			} 
		}
	}
	\Return{$V_r$}\;
	\caption{\label{algo:neighbor}{\sc \getrnd} for \algo.}
\end{algorithm}
\begin{figure*}[t]
	\vspace{-4.2mm} %
\end{figure*}

\subsection{Theoretical Analysis}
\label{sec:method:theory}
\vspace{-1mm}

We present a theoretical analysis of \algo (Alg.~\ref{algo:algo}).
We first prove the soundness and time complexity of \getrnd (Alg.~\ref{algo:neighbor}), which the performance of \algo relies on.
Then, we analyze the overall time complexity of \algo.
Lastly, we show that \algo requires sub-linear memory during the incremental summarization process.
{\bf All proofs are provided in Appendix~\ref{appendix:proofs}.}
 
\subsubsection{Soundness of \getrnd (Alg.~\ref{algo:neighbor})}
\label{sec:method:theory:rndnbd}

We show that \getrnd performs uniform sampling by proving Thms.~\ref{sec:method:theory:thm1} and \ref{sec:method:theory:thm2} that our solutions in Sect.~\ref{sec:method:algo} solve Problems~\ref{problem:neighbor} and \ref{problem:mcmc}. 

\begin{theorem}[Unbiasedness of our solution for Problem~\ref{problem:neighbor}]
\label{sec:method:theory:thm1}
	Our solution for Problem~\ref{problem:neighbor} draws a uniformly random node from $\bigcup_{i=1}^k N_i$.
	That is, each node is sampled with probability ${1}/{\sum_{i=1}^k |N_i|}$. 
\end{theorem}

\begin{theorem}[Soundness of our solution for Problem~\ref{problem:mcmc}]
\label{sec:method:theory:thm2}
	\sloppy
	Our solution for Problem~\ref{problem:mcmc}, which is an MCMC method equipped with (a) the proposal distribution $P(S_j | S_i) = 1/k$ for any $i, j$ and (b) the acceptance probability $A(S_j, S_i) = \textup{min}(1, {|S_j|}/{|S_i|})$ for any move from $S_i$ to $S_j$, asymptotically simulates $\pi$ in Eq.~\eqref{eq:probdist}.
\end{theorem}

\subsubsection{Time Complexity of \getrnd (Alg.~\ref{algo:neighbor})}
\label{sec:method:theory:time}

Before proving the time complexity of \getrnd, which is a core building block of \algo, 
we first prove the time complexity of retrieving the neighborhood of an input node from the outputs (i.e., $G^*$ and $C$) in Lemma~\ref{sec:method:theory:getnbd}.
We let $\Cp(u):=\{v \in V | \{u,v\} \in \Cp\}$ and $\Cm(u):=\{v \in V | \{u,v\} \in \Cm\}$ be the sets of neighbors of $u$ in $\Cp$ and $\Cm$, respectively, and we let $N(A) := \{B \in S | \{A, B\} \in P\}$ be the set of neighboring supernodes of the supernode $A$.

\begin{lemma}[Average time complexity of retrieving neighborhood]
	\label{sec:method:theory:getnbd}
	The average-case time complexity of retrieving the neighborhood of a node is $O(\overline{deg})$, where $\overline{deg}$ is the average degree.
\end{lemma}

As discussed in Limitation 2 in Sect.~\ref{sec:method:algobasic}, Lemma~\ref{sec:method:theory:getnbd} motivates us to design \getrnd, which is much faster than retrieving the entire neighborhood, as formulated in Thm.~\ref{sec:method:theory:rnd_time}.

\begin{theorem}[Average time complexity of \getrnd]
	\label{sec:method:theory:rnd_time}
	Assume the neighborhood in $\Cp$, $\Cm$ and $P$ of each node is stored in a hash table.
	The average-time complexity of Alg.~\ref{algo:neighbor} for each node $u$ is $O(c \cdot (1+{|\Cm(u)|}/{deg(u)}))$. %
\end{theorem}

The average time complexity of \getrnd called by \algo becomes constant,
under a realistic assumption \cite{albert2002statistical} that an initial graph evolves under $\stream$ where $e_t$ takes place on a node with probability proportional to the degree of the node. That is, higher-degree nodes are more likely to go through changes in their neighbors.
This result is shown in Corollary~\ref{sec:method:theory:rnd_final} based on Lemma~\ref{sec:method:theory:degree_lemma}.

\begin{lemma}
	\label{sec:method:theory:degree_lemma}
	Let $X$ be a discrete random variable whose domain is $\{\frac{a_u}{b_u}\}_{u \in V}$, where $a_u = |\Cm(u)|$ and $b_u = deg(u)$ for each node $u$ and the probability mass is proportional to $deg(u)$.
	Then, $E[X] \leq 1$.
\end{lemma}

\begin{corollary}[Average time complexity of \getrnd under preferential attachment \cite{albert2002statistical}]
	\label{sec:method:theory:rnd_final}
	The average-case time complexity of \getrnd called by Alg.~\ref{algo:algo} is $O(c)$ under the assumption that each edge change $e_t$ takes place on nodes with probability proportional to the degree of the nodes.
\end{corollary}

\subsubsection{Overall Time Complexity of \algo (Alg.~\ref{algo:algo})}
\label{sec:method:theory:overall}

In our analysis below,
we let $SN(A) := \{B \in S : \exists (u, v) \in A \times B, \textup{ s.t. } \{u , v\}\in E \}$ be the set of supernodes connected to $A$ by at least one edge.
We also let $E_{AB} := \{ \edge \in E | u\in A, v \in B, u\neq v\}$ and $T_{AB} := \{ \edge | u\in A, v \in B, u\neq v\}$ be the sets of existing and potential edges between supernodes $A$ and $B$, respectively.

\smallsection{Computing Savings in the Objective $\varphi$ (Eq.~\eqref{eq:objective}):}
Suppose that we move a node $y$ from a supernode $S_{y}$ to a supernode $S_{z}$. Then, the previous optimal encoding (see Sect.~\ref{sec:method:term}) between $S_{y}$ and each of $SN(S_y)$ and between $S_{z}$ and each of $SN(S_z)$ can be affected, while that between the other supernodes should remain the same.  Hence, only $SN(S_{y})$ and $SN(S_{z})$ are of concern when computing the saving in the objective $\varphi$ (i.e., Eq~\eqref{eq:objective}) induced by the move of $y$. Thus, the time complexity of this task is $O(|SN(S_{y})| + |SN(S_{z})|)$.

\smallsection{Updating Optimal Encoding:}
The computation of the saving determines whether a move is accepted or not. 
Once the move is accepted, the previous optimal encoding should be updated in response to the move.
As explained in Sect.~\ref{sec:method:term}, the edges between two supernodes A and B are optimally encoded either as (1) $|E_{AB}|$ edges in $\Cp$ or as (2) the superedge $\{A, B\}$ with $|T_{AB}| - |E_{AB}|$ edges in $\Cm$. 
In the latter case, $|T_{AB}| - |E_{AB}|+ 1 \leq |E_{AB}|$ holds in the optimal encoding. 
Hence, in both cases, the number of references needed to update the optimal encoding between each supernode pair $\{A, B\}$ is $O(|E_{AB}|)$.
Due to the reason discussed in the previous paragraph, the total worst-case time complexity is
\begin{equation*}
	O\Big(\sum_{U \in SN(S_{y})} \mathclap{\ |E_{S_{y}U}|} \quad \ \ + \sum_{U \in SN(S_{z})} \mathclap{\ |E_{S_{r}U}|} \quad \ \ \  \Big)
	= O\Big(\sum_{u \in S_{y}} deg(u) + \sum_{u \in S_{z}} deg(u)\Big). \vspace{-1mm}
\end{equation*}

\smallsection{Scalability in Practice:} 
In practice, the above two subtasks do not harm the scalability of \algo, which proceeds to trials only with probability $\frac{1}{deg}$ and updates the optimal encoding only if a move is accepted.
We show in Appendix.~\ref{appendix:exp:parameter} that
the runtime of \algo in practice increases proportionally to the number of samples (i.e., $c$) per input change, as expected from Corollary~\ref{sec:method:theory:rnd_final}.
For a fixed $c$, \algo processes each change in {\bf near-constant time}, regardless of the growth of the input graph, as we show empirically in Sect.~\ref{sec:exp:scalability}.

\begin{table}[t]
	\centering
	\small
	\caption{\label{tab:dataset} 
		Real-world graph streams used in our experiments. %
	}
	\scalebox{0.88}{
		\hspace{-2.5mm}
		\begin{tabular}{l|l|l|l|l}
			\toprule
			Name & \#Nodes & \#Insertions & \#Deletions & Summary  \\ 
			\midrule
			\multicolumn{4}{l}{Insertion-only (IO) graph streams} \\
			\midrule
			Protein (PR) & $6,229$ & $146,160$ & - & Protein Interaction \\ %
			Email-Enron (EN) & $86,977$ & $297,456$ & - & Email \\
			Facebook (FB) & $61,095$ & $614,797$ & - & Friendship \\
			Web-EU-05 (EU)  & $862,664$ & $16,138,468$ & - & Hyperlinks \\ %
			Hollywood (HW) & $1,985,306$ & $114,492,816$ & - & Collaboration \\ %
			Web-UK-02 (UK) & $18,483,186$ & $261,787,258$ & - & Hyperlinks \\ %
			\midrule
			\multicolumn{4}{l}{Fully-dynamic (FD) graph streams} \\
			\midrule
			DBLP (DB) & $317,080$ & $1,049,866$ & $116,042$ & Coauthorship \\
			YouTube (YT) & $1,134,890$ & $2,987,624$ & $331,305$ & Friendship \\
			Skitter (SK) & $1,696,415$ & $11,095,298$ & $1,233,226$ & Internet \\
			LiveJournal (LJ) & $3,997,962$ & $34,681,189$ & $3,854,423$ & Friendship \\
			\bottomrule
	\end{tabular}}
\end{table}	

\subsubsection{Space Complexity of \algo (Alg.~\ref{algo:algo})}
\label{sec:method:theory:space}

The sub-linear space complexity of \algo proved in Thm.~\ref{sec:method:theory:thm_space} implies that the entire input graph of size $O(|V|+|E|)$ does not have to be maintained in memory during the incremental summarization process.

\vspace{-1mm}
\begin{theorem}[Space complexity of \algo]
	\label{sec:method:theory:thm_space}
	The space complexity of Algorithm \ref{algo:algo} is $O(|V|+|P|+|\Cp|+|\Cm|)$.  
\end{theorem}
\vspace{-1mm}

\vspace{-1mm}
\section{Experiments}
\vspace{-1mm}
\label{sec:exp}
\begin{figure*}[t]
	\centering
	\vspace{-4.5mm}
	\includegraphics[width=1.8\columnwidth]{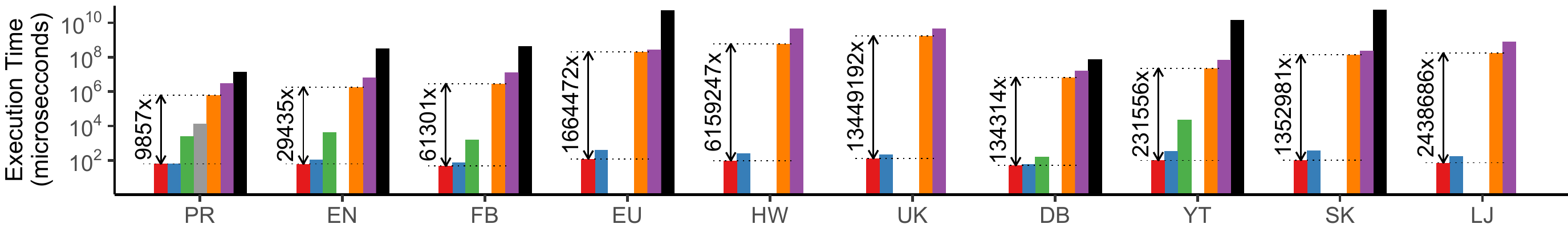}
	\includegraphics[width=0.26\columnwidth]{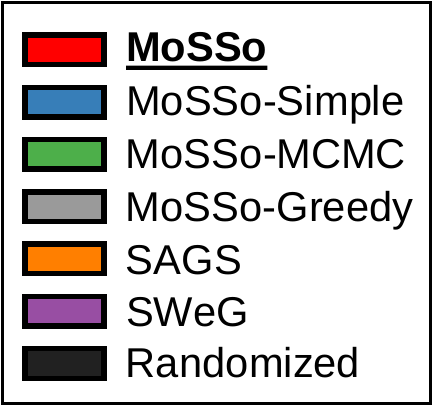}
	\caption{\label{fig:q1} \underline{\smash{\algo is fast.}} It processed each change in the input graph up to {\bf 7 orders and 2 orders} of magnitude faster than the fastest batch algorithm and the streaming baselines (\sgreedy and \mcmc), respectively. The baseline streaming algorithms and \rnd ran out of time ($>$ 24 hours) on some large graphs and do not appear in the corresponding plots.}
\end{figure*}

We review our experiments for answering the following questions:
\begin{enumerate}[leftmargin=*]
	\item[Q1.] \textbf{Speed:} How fast is \algo, compared to the baseline streaming algorithms and the best batch algorithms?
	\item[Q2.] \textbf{Compression Ratio:} How compact are the outputs of \algo, compared to those obtained by the best competitors?
	\item[Q3.] \textbf{Scalability:} How does the runtime of \algo grow as the input graph grows?
	\item[Q4.] \textbf{Performance Analysis (Appendix~\ref{appendix:exp}):} How do graph properties and parameter values affect the performance of \algo?
\end{enumerate}

\vspace{-1mm}
\subsection{Experimental Settings}
\label{sec:exp:settings}
\vspace{-1mm}

\smallsection{Machine:} We performed our experiments on a desktop with a 3.7GHz Intel i5-9600k CPU and 64GB memory.

\smallsection{Dataset:} We used ten real-world graphs listed in Table~\ref{tab:dataset}. 
In every graph, we ignored the direction of all edges and removed both self-loops and multiple edges. 
From the graphs, we generate both insertion-only and fully dynamic graph streams with deletions. %
We generated an insertion-only graph stream by listing the insertions of all edges in a graph. We either sorted the insertions by their timestamps (if they exist) or randomly ordered them (otherwise). 
For a fully-dynamic graph stream, we listed the insertions of all edges in a graph after randomly ordering them.
Then, for each edge, we created the deletion of it with probability $0.1$ and located the deletion in a random position after the insertion of the edge.

\smallsection{Implementation:} We implemented the following lossless graph-summarization algorithms in OpenJDK 12:
(a) \algo %
 (\textbf{proposed}) with $e$=$0.3$, $c$=$120$;
(b) \algobasic (\textbf{proposed}) with $e$=$0.3$, $c$=$120$;
(c) \sgreedy (streaming baseline),
(d) \mcmc (streaming baseline) with $\beta = 10$;
(e) \sweg \cite{shin2019sweg} (batch baseline) with $\#$Threads=$1$, $T$=$20$,  $\epsilon$=$0$;
(f) \sags \cite{khan2015set} (batch baseline) with $h$=$30$, $b$=$10$, $p$=$0.3$;
and (g) \rnd \cite{navlakha2008graph} (batch baseline).

\smallsection{Evaluation Metric:} Given a summary graph $G^{*} = \summary$ and edge corrections $C = \cor$ of a graph $G = (V, E)$, we used
\vspace{-1mm}
\begin{equation}
	({|P|+|C^+|+|C^-|})/{|E|} \label{eq:evalmetric} \vspace{-1mm}
\end{equation}
as the \textit{\ratio}.
In Eq.~\eqref{eq:evalmetric}, the numerator is the objective $\varphi$, and the denominator is a constant, given an input graph $G$.
Eq.~\eqref{eq:evalmetric} and runtime were averaged over $3$ trials unless otherwise stated.

\vspace{-1mm}
\subsection{Q1. Speed (Fig.~\ref{fig:q1})}
\label{sec:exp:speed}
\vspace{-1mm}

To evaluate the speed of \algo, we measured the runtime of all considered algorithms.
Specifically, for batch algorithms, we measured the time taken for summarizing each dataset.
For streaming algorithms, we measured time taken for processing all the changes in each dataset, and then we divided it by the number of changes to compute the time taken for processing each change.

\textbf{\algo was significantly faster than its competitors.}
As seen in Fig.~\ref{fig:q1}, it processed each change up to \underline{\smash{7 orders of magnitude}} \underline{\smash{faster}} than running the fastest batch algorithm. 
The gap between \algo and the batch algorithms was wide in large datasets. 
Moreover, \algo was up to \underline{\smash{2 orders of magnitude faster}} than the baseline streaming algorithms, and the gap increased in larger datasets, where the baseline method ran out of time.
The two versions of \algo showed comparable speed in small datasets, while \algo was faster than \algobasic in large datasets.

\begin{figure*}[t]
	\vspace{-6.5mm}
	\centering
	\subfigure[Protein (IO)]{
		\includegraphics[width=.35\columnwidth]{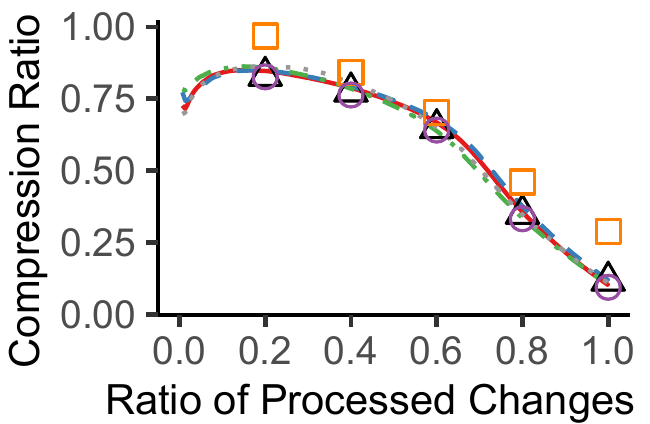}
	}\hspace{-2mm}
	\subfigure[Email-Enron (IO)]{
		\includegraphics[width=.35\columnwidth]{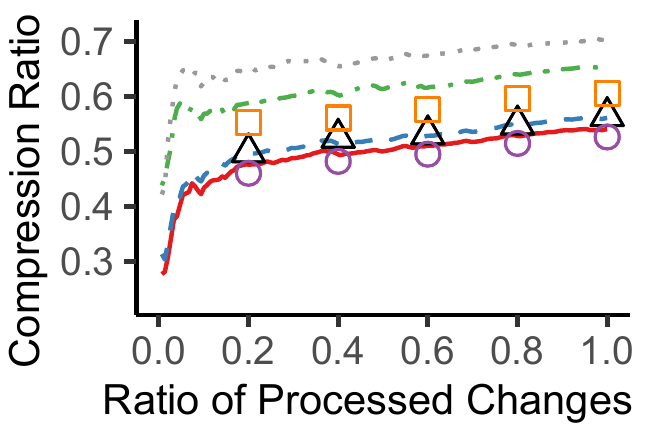}
	}\hspace{-2mm}
	\subfigure[Facebook (IO)]{
		\includegraphics[width=.35\columnwidth]{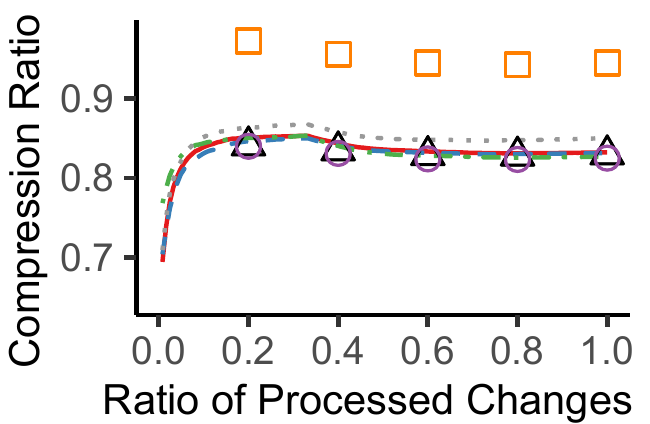}
	}\hspace{-2mm}
	\subfigure[DBLP (FD)]{
		\includegraphics[width=.35\columnwidth]{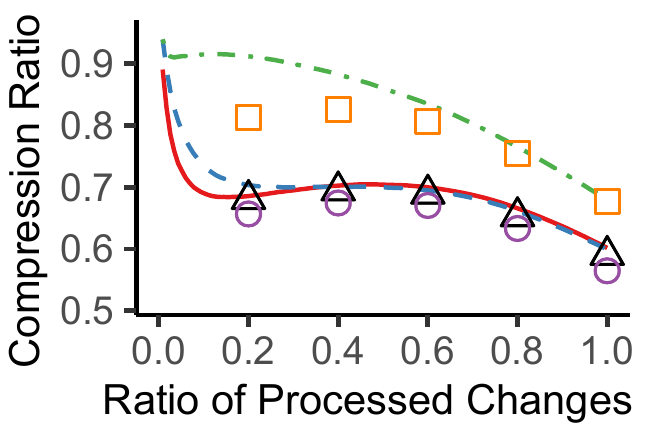}
	}\hspace{-2mm}
	\subfigure[YouTube (FD)]{
		\includegraphics[width=.35\columnwidth]{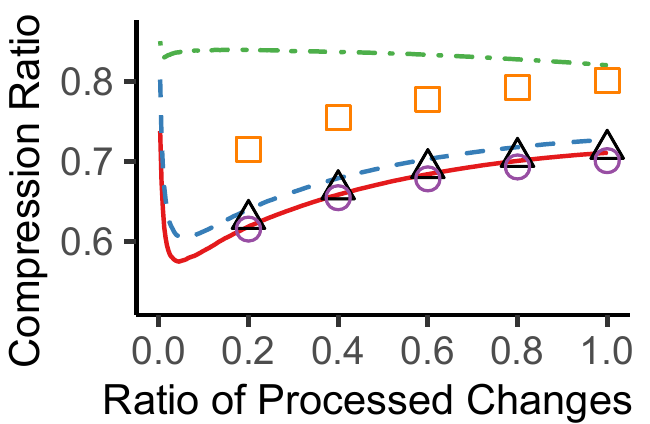}
	}\includegraphics[width=.27\columnwidth]{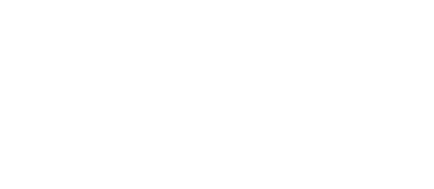}
	\\ \vspace{-3mm}
	\subfigure[Web-EU-05 (IO)]{
		\includegraphics[width=.35\columnwidth]{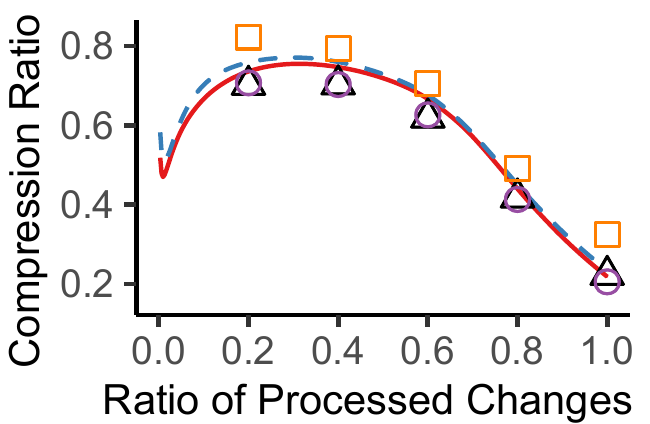}
	}\hspace{-2mm}
	\subfigure[Hollywood (IO)]{
		\includegraphics[width=.35\columnwidth]{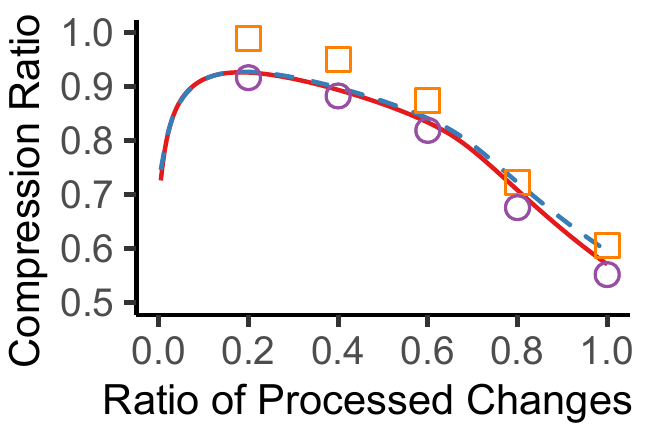}
	}\hspace{-2mm}
	\subfigure[Web-UK-02 (IO)]{
		\includegraphics[width=.35\columnwidth]{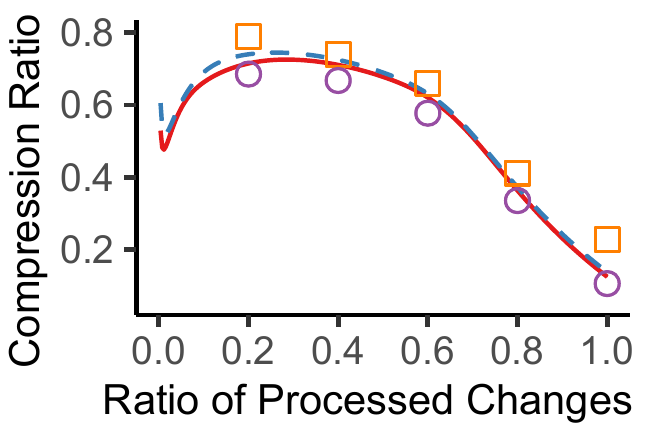}
	}\hspace{-2mm}
	\subfigure[Skitter (FD)]{
		\includegraphics[width=.35\columnwidth]{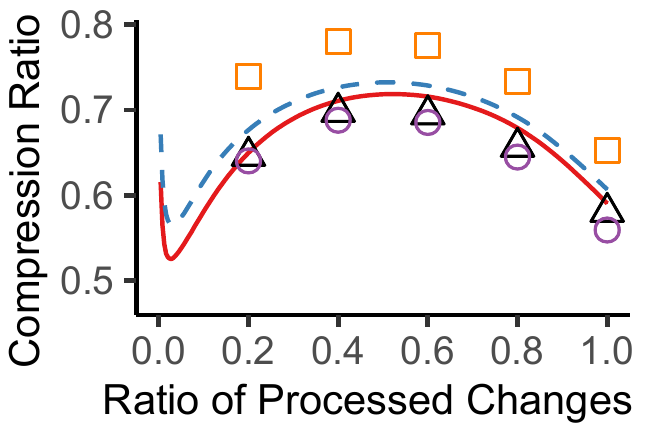}
	}\hspace{-2mm}
	\subfigure[LiveJournal (FD)]{
		\includegraphics[width=.35\columnwidth]{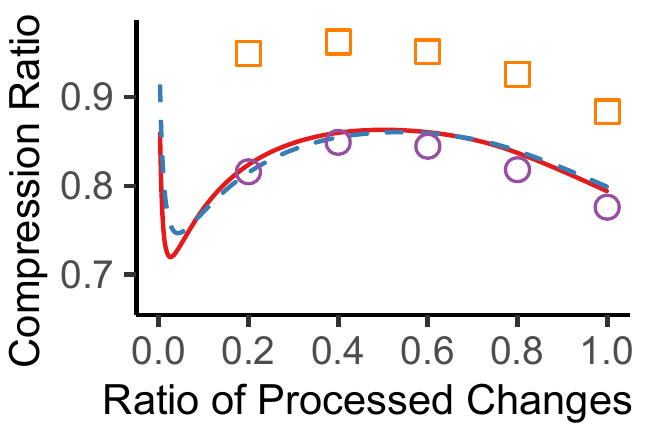}
	}
	\includegraphics[width=.26\columnwidth]{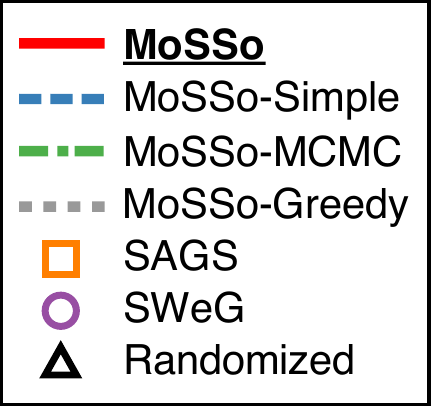}
	\\
	\vspace{-2mm}
	\caption{\label{fig:q2} 
		\underline{\smash{\algo is `any time' with compact representations.}}
		\algo achieved the best \ratios among streaming algorithms, which always maintain the summarized representation of the input graph.
		Moreover, \algo achieved \ratios comparable to that of the best batch algorithms, which were rerun repeatedly. The baseline streaming algorithms and \rnd ran out of time ($>$ 24 hours) on some large graphs and so, do not appear in the corresponding plots. %
	}
\end{figure*}

\vspace{-1mm}
\subsection{Q2. Compression Ratio (Fig.~\ref{fig:q2})}
\label{sec:exp:compress}
\vspace{-1mm}

To test the effectiveness of \algo,
we measured the compactness of outputs obtained by all considered algorithms using the \ratio, defined in Eq~\eqref{eq:evalmetric}.
For streaming algorithms, 
we tracked the changes in the \ratio while each input graph stream evolved over time.
For batch algorithms, we ran them when $20\%$, $40\%$, $60\%$, $80\%$ and $100\%$ of the changes in each dataset arrived. 

\textbf{\algo achieved the best \ratios among the streaming algorithms}, as seen in Fig.~\ref{fig:q2}.
Moreover, the \ratio of \algo was even comparable to those of the best batch algorithms.
Between the two versions of \algo, \algo consistently showed better \ratios than \algobasic.

\vspace{-1mm}
\subsection{Q3. Scalability (Fig.~\ref{fig:ad:scalable})}
\label{sec:exp:scalability}
\vspace{-1mm}

To evaluate the scalability of \algo, we measured its runtime in the Web-EU-05 dataset. 
As seen in Fig. \ref{fig:ad:scalable}, the accumulated runtime of \algo was near-linear in the number of processed changes, implying that {\bf \algo processed each change in near-constant time}.
Note that we ignored time taken to wait for input changes to arrive.
More extensive results can be found at Appendix~\ref{appendix:scalability}.

\vspace{-1mm}
\section{Related Works}
\vspace{-1mm}
\label{sec:related}

Graph summarization has been studied in a number of settings, and an extensive survey can be found in \cite{liu2018graph}.
We focus on studies directly related to our work: (1) lossless summarization of static graphs, %
and (2) lossy and lossless summarization of dynamic graphs.

\smallsection{Lossless summarization of static graphs:} Lossless summarization algorithms on static graphs output a summary graph with edge corrections. \greedy \cite{navlakha2008graph} examines all pairs of supernodes that are 2-hops away, and among the pairs, it repeatedly merges the pair leading to the largest saving in the objective $\varphi$. 
\rnd \cite{navlakha2008graph} %
speeds up the process by randomly picking one supernode and searching only for the second supernode that leads to the largest saving in $\varphi$, when being merged with the first one.
To further reduce the time complexity, \sags \cite{khan2015set} selects two supernodes to be merged based on locality sensitive hashing.
While these algorithms fail to strike a balance between compression and time complexity, a parallel algorithm \sweg \cite{shin2019sweg} succeeds in improving both parts.
Note that these batch algorithms are not designed to address changes in the input graph and should be rerun from scratch to reflect such changes. 
Lossy variants of the graph summarization problem were also explored in \cite{lefevre2010grass,riondato2017graph,navlakha2008graph,shin2019sweg,lee2020ssumm}.

\smallsection{Summarization of dynamic graphs:} 
Since the dynamic nature of graphs makes lossless summarization more challenging, previous studies of dynamic graph summarization have focused largely on lossy summarization for query efficiency \cite{tang2016graph, gou2019}, social influence analysis \cite{mehmood2013csi, mathioudakis2011sparsification}, temporal pattern \cite{tsalouchidou2016scalable}, etc.
They aim to summarize: either the current snapshot of the input graph \cite{zhao2011gsketch,tang2016graph, khan2017toward, gou2019} or temporal patterns in the entire history of the growth of the input graph \cite{tsalouchidou2016scalable}.
For the former, existing incremental algorithms \cite{zhao2011gsketch,tang2016graph, khan2017toward, gou2019} produce lossy summary for query efficiency in a form of graph sketch (i.e., an adjacency matrix smaller than the original one).
Especially, \textsc{TCM} \cite{tang2016graph} and its variant \textsc{GSS} \cite{gou2019} compress the original graph $G=(V,E)$ into a graph sketch $G_s$ by using a hash function $h(\cdot )$ through mapping a node $v\in V$ into a node $h(v)$ in $G_v$ and mapping an edge $\{u, v\}\in E$ into $\{h(u), h(v)\}$ in $G_v$.
For the latter, $kC$ \& $k\mu$ \cite{tsalouchidou2016scalable} lossily compress the entire history of a dynamic graph to find temporal patterns. %
While \textsc{TimeCrunch} \cite{shah2015timecrunch} is lossless, it is a batch algorithm for summarizing the entire growth history of a dynamic graph.
To the best of our knowledge, \algo is the first incremental algorithm for summarizing the current snapshot of a fully dynamic graph in a lossless manner.

\vspace{-1mm}
\section{Conclusions}
\vspace{-1mm}
\label{sec:conclusions}

We propose \algo, a simple but fast and effective incremental algorithm for lossless summarization of fully-dynamic graphs. 
\algo is based on several novel ideas, including \alone, \fastr, and \guide, and
they lead to significant improvement in speed and compression rates.
We empirically and theoretically show that \algo has the following strengths:
\vspace{-0.5mm}
\begin{itemize}[leftmargin=*]
  \item {\bf Fast and `any time'}: In response to each change, \algo updates its lossless summary up to {\bf 7 orders of magnitude faster} than the fastest batch algorithm (Fig.~\ref{fig:q1}).
  The update time remains constant while the input graph grows (Fig.~\ref{fig:ad:scalable}).
  \item {\bf Scalable}: \algo successfully summarizes a fully dynamic graph with up to $0.3$ billion edges, without having to maintain the original graph in memory (Thm.~\ref{sec:method:theory:thm_space}).
  \item {\bf Effective}: \algo shows compression rates comparable to state-of-the-art batch algorithms (Fig.~\ref{fig:q2}).
\end{itemize}
\vspace{-0.5mm}
{\bf Reproducibility}: The code and datasets used in the paper are available at \url{http://dmlab.kaist.ac.kr/mosso/}.

\begin{figure*}[t]
	\centering
	\vspace{-7mm}
	\subfigure[\label{fig:q4a} Effects of the escape probability (i.e., $e$)]{
		\includegraphics[width=.396\columnwidth]{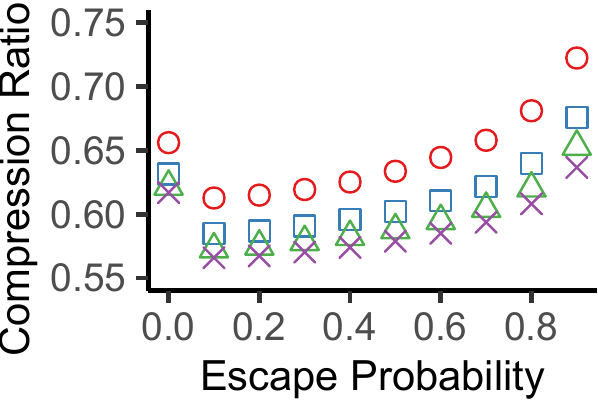}
		\includegraphics[width=.44\columnwidth]{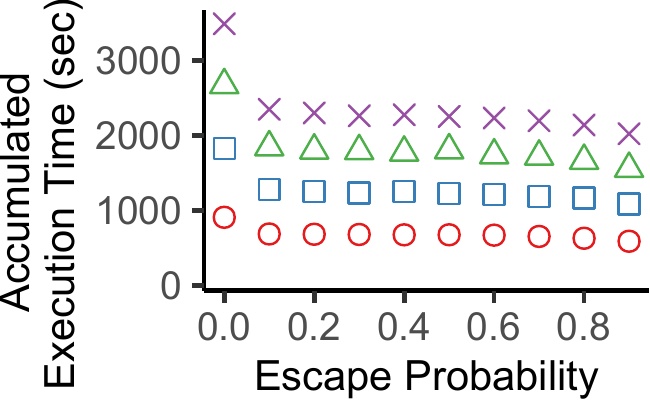}
		\includegraphics[width=.16\columnwidth]{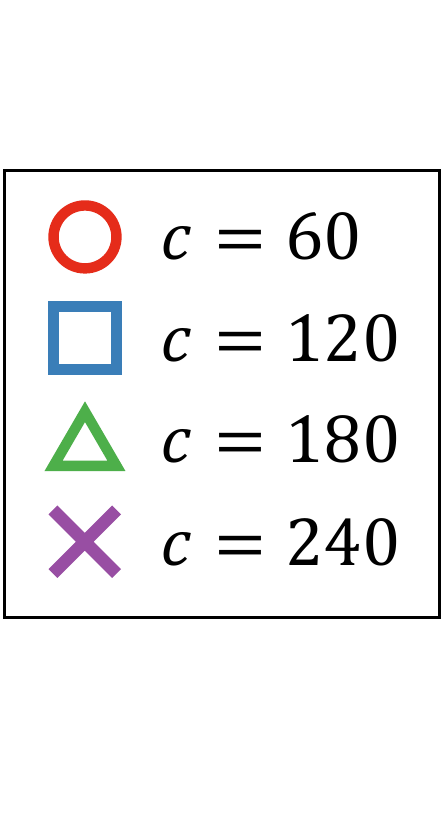}
	}
	\subfigure[\label{fig:q4b} Effects of the number of samples (i.e., $c$)]{
		\includegraphics[width=.396\columnwidth]{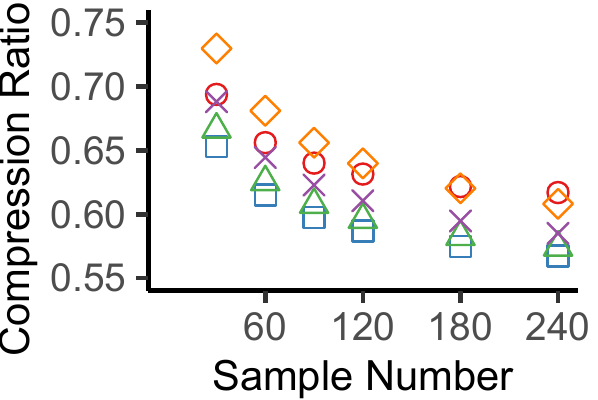}
		\includegraphics[width=.44\columnwidth]{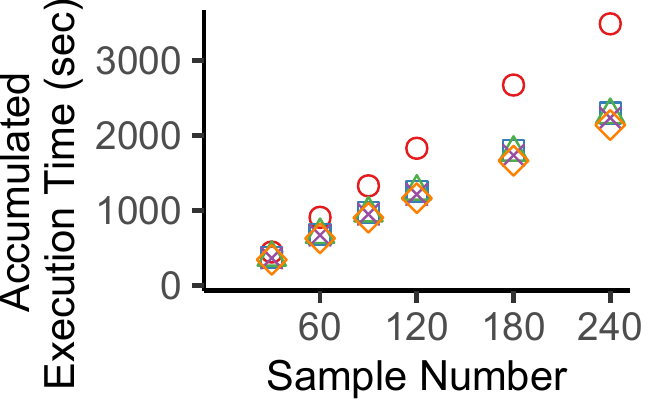}
		\includegraphics[width=.16\columnwidth]{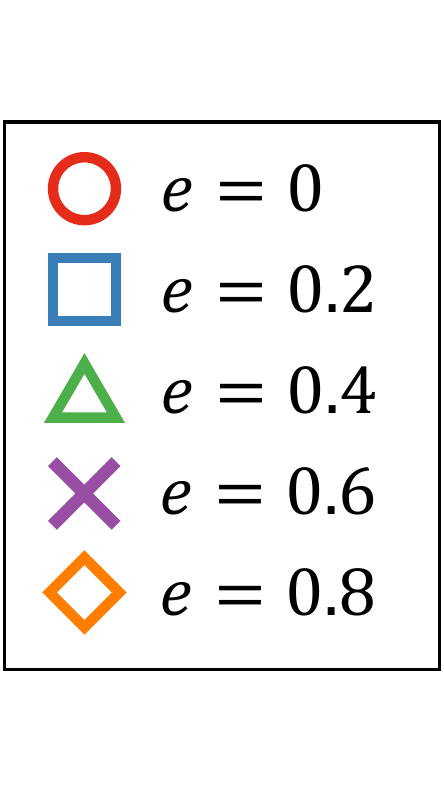}
	} \\
	\vspace{-2mm}
	\caption{\label{fig:q4} 
		\underline{\smash{Effects of the parameters of \algo.}}
		(a) \algo gave the most compact output when the escape probability $e$ was around $0.1$. As $e$ increased, \algo tended to yield larger outputs without much change in its execution time. (b) As the number of samples $c$ increased, \algo tended to take more time with more compact outputs.}
\end{figure*}

\begin{figure*}[t]
	\vspace{-3mm}
	\centering
	\subfigure[\label{fig:appendix:copying} Effect of copying probability]{
		\includegraphics[width=.35\columnwidth]{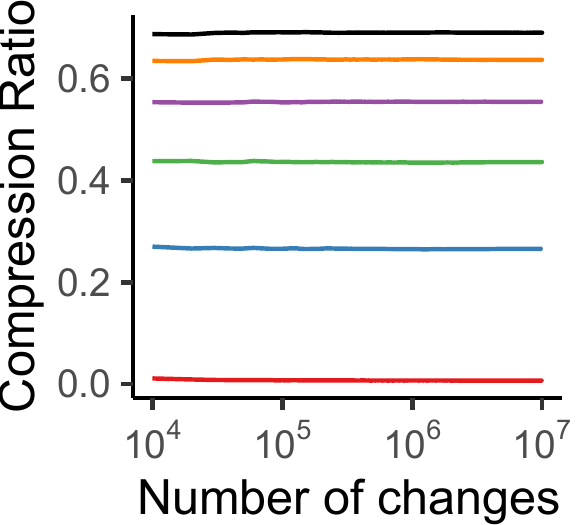} \hspace{1mm}
		\raisebox{0.10\height}{\includegraphics[width=.18\columnwidth]{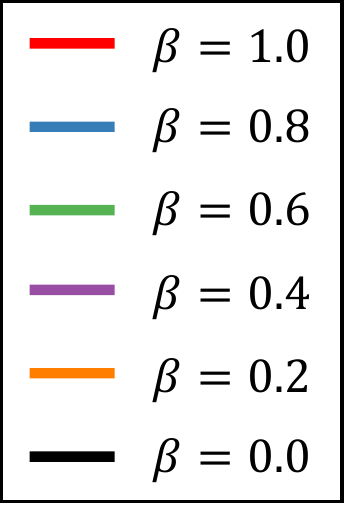}}
	}
	\subfigure[\label{fig:appendix:scalability1} Scalability of \algo and \algobasic (WEB-EU-05)]{
		\includegraphics[width=.35\columnwidth]{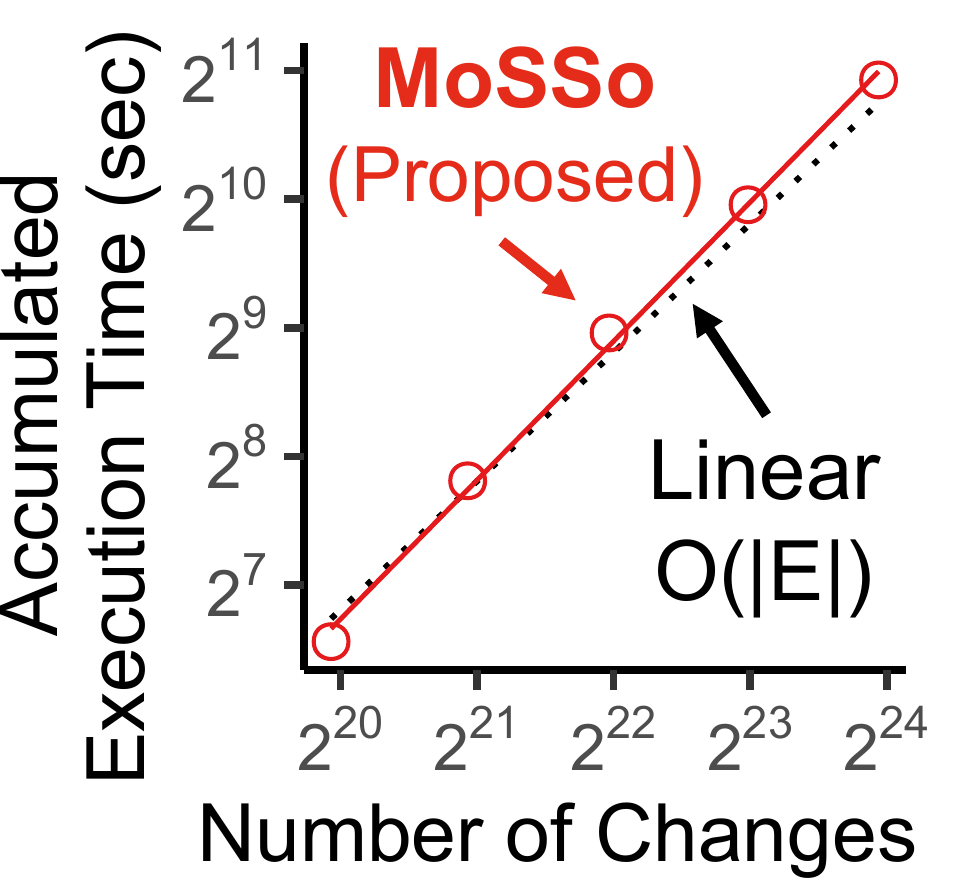}
		\includegraphics[width=.35\columnwidth]{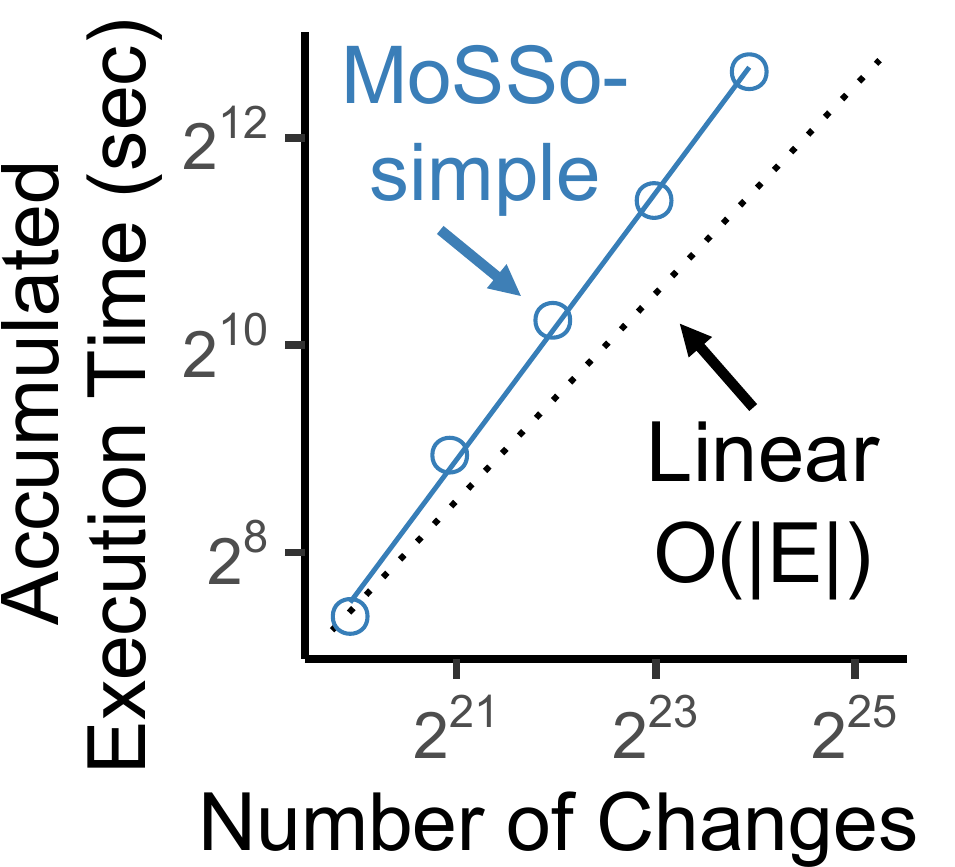}
	} 
	\subfigure[\label{fig:appendix:scalability2} Scalability of \algo and \algobasic (Skitter)]{
		\includegraphics[width=.35\columnwidth]{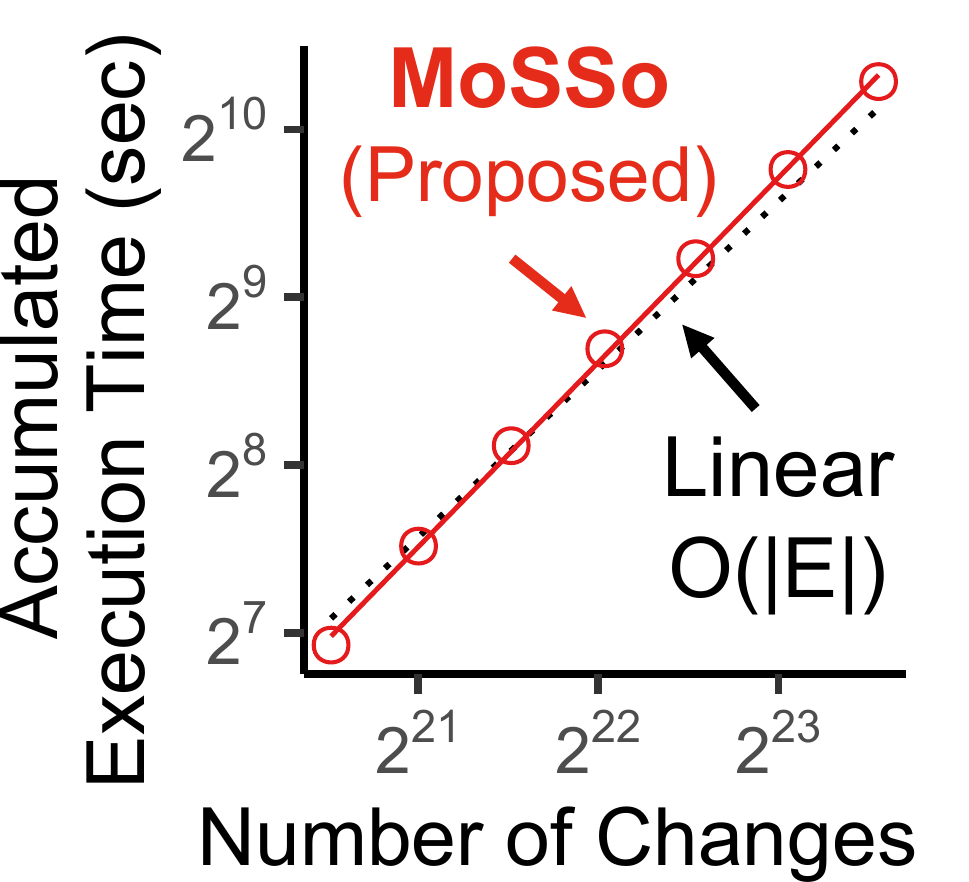}
		\includegraphics[width=.35\columnwidth]{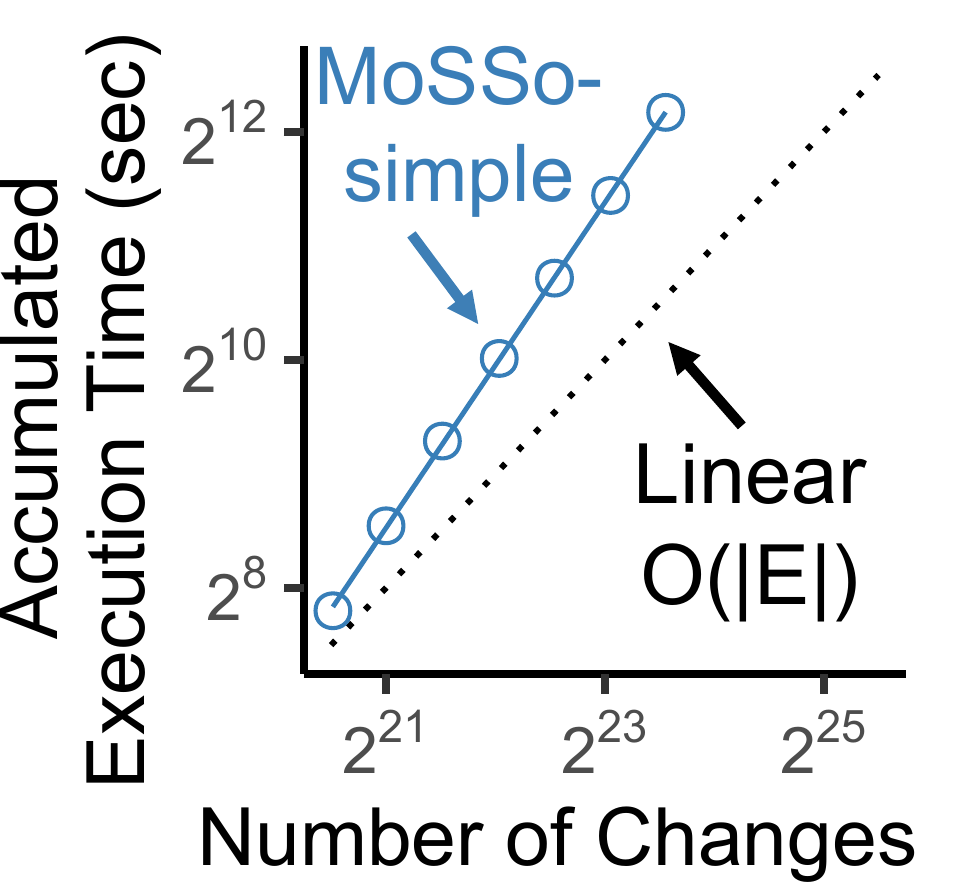}
	} \\
	\vspace{-2mm}
	\caption{
		(a) \underline{\smash{Effects of input graph properties.}} More nodes with similar connectivity (i.e., higher copying probability $\beta$) led to better compression. 
		(b,c) \underline{\smash{\algo is more scalable than \algobasic.}} \algo took near-constant time per change (i.e., accumulated runtime $\propto |E|$), while \algobasic took more over time (i.e., accumulated runtime $\propto |E|^{1.42}$).}
\end{figure*}

\vspace{-1mm}
\subsection*{Acknowledgements}
\vspace{-1mm}
{\small This work was supported by National Research Foundation of Korea (NRF) grant funded by the
Korea government (MSIT) (No. NRF-2019R1F1A1059755) and Institute of Information \& Communications
Technology Planning \& Evaluation (IITP) grant funded by the Korea government (MSIT) (No. 2019-0-00075, Artificial Intelligence Graduate School Program (KAIST)).}

\balance

\vspace{-1.5mm}
\bibliographystyle{ACM-Reference-Format}
\bibliography{BIB/kijung,BIB/yunbum,BIB/jihoon}

\newpage
\appendix
\label{appendix}

\appendix

\section{Appendix: Additional Experiments}
\label{appendix:exp}

\vspace{-1mm}
\subsection{Effects of Parameters (Fig.~\ref{fig:q4})}
\label{appendix:exp:parameter}
\vspace{-1mm}

We analyzed the effects of the parameters of \algo on its speed and \ratio.
To this end, we measured how its accumulated runtime and \ratio changed depending on the number $c$ of samples and the escape probability $e$ in the Skitter dataset.

As seen in Fig.~\ref{fig:q4}, {\bf the \ratio was minimized when the escape probability $e$ was around $0.1$}, and the \ratio tended to increase as $e$ increased.
While the runtime decreased as $e$ increased, the amount of change was small.
As the number of samples $c$ increased, the \ratio decreased gradually, while the runtime increased proportionally to $c$.

\vspace{-1mm}
\subsection{Effects of Graph Properties (Fig.~\ref{fig:appendix:copying})}
\vspace{-1mm}
To analyze the effect of graph properties on the compression rates of \algo, we used the copying model \cite{kleinberg1999web} with varying copying probabilities.
Specifically, we created synthetic graph with $|V| = |E| = 10,000,000$ based on the model\footnote{We symmetrized output graphs and removed all self-loops.} and measured compression rates for each copying probability $\beta$. As seen in Fig.~\ref{fig:appendix:copying}, more nodes with similar connectivity (i.e., higher copying probability $\beta$) led to better compression.

\vspace{-1mm}
\subsection{Additional Scalability Tests (Figs.~\ref{fig:appendix:scalability1}-\ref{fig:appendix:scalability2})}
\label{appendix:scalability}
\vspace{-1mm}
We measured accumulated runtime of \algo and \algobasic in the Web-EU-05 and Skitter datasets. As seen in Figs. \ref{fig:appendix:scalability1}-\ref{fig:appendix:scalability2}, the accumulated runtime of \algo was near-linear in both datasets. However, the accumulated runtime of \algobasic was super-linear (proportional to $|E|^{1.42}$).

\vspace{-1mm}
\section{Appendix: Proofs}
\label{appendix:proofs}
\begin{proofof}{Thm.~\ref{sec:method:theory:thm1}}
	In our solution, we sample twice in a row: we first draw a supernode according to Eq.~\eqref{eq:probdist} and then draw a node in the supernode. 
	For any given node $u$ in $\bigcup_{i=1}^k S_i$, the probability of drawing $u$ is
	
	\vspace{-4mm}
	\begin{align*}
	P& (u  \textup{ is selected}) \\
	& = P(S_u \textup{ is selected}) \cdot P(u \textup{ is selected} | S_u \textup{ is selected})\\
	& = \frac{|S_u|}{|S_1| + \cdots + |S_k|} \cdot \frac{1}{|S_u|} = \frac{1}{|S_1|+ \cdots + |S_k|}
	= \frac{1}{S},
	\end{align*}
	
	\noindent where $S := |\bigcup_{i=1}^k S_i| = \sum_{i=1}^k|S_i|$. Hence, each node in $\bigcup_{i=1}^k S_i$ is equally likely to be drawn under this scheme.
	Thus, a node not in $\bigcup_{i=1}^k N_i$ is drawn with probability $\frac{S-N}{S}$. In such a case, we fail and repeat the process from beginning. 
	Then, the probability of sampling each node $w \in \bigcup_{i=1}^k N_i$ is 
	\begin{align*}
	&P(w \textup{ is selected}) = \sum\nolimits_{i=1}^{\infty} P(w \textup{ is drawn at the }i\textup{-th trial})\\
	&= \sum\nolimits_{i=1}^{\infty} P(w \textup{ is drawn at the }i\textup{-th trial}) \cdot P(\textup{fail } (i-1) \textup{ times})\\
	&= \sum\nolimits_{i=1}^{\infty} \frac{1}{S} \cdot \left(\frac{S-N}{S}\right)^{i-1} = \frac{1}{S} \cdot \frac{1}{1-\frac{S-N}{S}} = \frac{1}{N}. \qedhere
	\end{align*}
\end{proofof}

\noindent\begin{proofof}{Thm.~\ref{sec:method:theory:thm2}}
	To ensure that a sampling scheme converges to a target distribution, the sampling process is set to be ergodic (i.e. any supernode can be proposed regardless of the currently sampled supernode) and to satisfy detailed balance (i.e. $\pi_i P_{ij} = \pi_j P_{ji}$). Hence, for ergodicity, we set the proposal probability distribution in the MCMC method to $P_{ij} = P(S_j|S_i) = \frac{1}{k}$ for any $i, j$. That is, the probability that $S_j$ is selected as the next supernode when the current supernode is $S_i$ is $P_{ij}$. Even though $\pi$ and $P$ do not fulfill detailed balance, from the results in \cite{hastings1970monte}, introducing the acceptance probability $A(S_j, S_i) = \textup{min}(1, \frac{\pi_j}{\pi_i} \frac{P_{ji}}{P_{ij}}) = \textup{min}(1, \frac{|S_j|}{|S_i|})$ with which one accepts each proposed sample guarantees that the sampling scheme asymptotically converges to $\pi$.	
\end{proofof}

\noindent \begin{proofof}{Lem.~\ref{sec:method:theory:getnbd}}
	In order to retrieving the neighborhood $N(u)$ of a node $u$, we need to look up all nodes in any supernode in $N(S_u)$ and those in $\Cp(u)$ and $\Cm(u)$.
	Thus, it takes $O(\sum_{S_v \in N(S_u)}|S_v| + |\Cm(u)| + |\Cp(u)|)$. Since $\sum_{S_v \in N(S_u)} |S_v| + |\Cp(u)| = deg(u) + |\Cm(u)|$, the time complexity becomes $O(deg(u) + 2|\Cm(u)|)$.
	Let $a_u = |\Cm(u)|$ and $b_u = deg(u)$. Then, the average-case time complexity of retrieving the neighborhood of a node is:
	\begin{equation*}
		\frac{\sum_{u \in V} (2a_u + b_u)}{|V|} = \frac{2\sum_{u\in V} a_u + \sum_{u \in V} b_u}{|V|} = \frac{4|\Cm| + 2|E|}{|V|},
	\end{equation*}
	where the second equality comes from $\sum_{u\in V} |\Cm(u)|$ $= 2|\Cm|$ and $\sum_{u\in V} deg(u) = 2|E|$. By the optimal encoding (Sect.~\ref{sec:method:term}), $|P|+|\Cp|+|\Cm| \leq |E|$ holds and thus, $|\Cm| \leq |E|$. Using this inequality, the average-case time complexity is bounded by $3\cdot\overline{deg}$.
\end{proofof}

\noindent \begin{proofof}{Thm.~\ref{sec:method:theory:rnd_time}}
	A node is drawn from $\Cp(u)$ with probability $\frac{|\Cp(u)|}{deg(u)}$ and from $\bigcup_{i=1}^k S_i$ with probability $\frac{deg(u)-|\Cp(u)|}{deg(u)}$.
	The former takes $O(1)$, but the latter requires repeated sampling until a node in $\bigcup_{i=1}^k S_i$ is drawn.
	As in the proof of Thm.~\ref{sec:method:theory:thm1}, a neighbor of $u$ in $\bigcup_{i=1}^k N_i$ is drawn with probability $\frac{N}{S}$ in each trial.
	Thus, the expected number of trials is simply $\frac{S}{N}$, and the average-case time complexity of \getrnd for sampling a node is
	
	\begin{align*}
	&\frac{|\Cp(u)|}{deg(u)} + \frac{deg(u) - |\Cp(u)|}{deg(u)} \cdot \frac{S}{N}\\
	&= \frac{|\Cp(u)|}{deg(u)} + \frac{deg(u) - |\Cp(u)|}{deg(u)} \cdot \frac{deg(u) - |\Cp(u)| + |\Cm(u)|}{deg(u) - |\Cp(u)|}\\
	&= \frac{|\Cp(u)|}{deg(u)} + \frac{deg(u) - |\Cp(u)| + |\Cm(u)|}{deg(u)} = 1 + \frac{|\Cm(u)|}{deg(u)}.
	\end{align*}
	Hence, the average-case time complexity of \getrnd for $c$ samples is $O(c \cdot (1 + \frac{|\Cm(u)|}{deg(u)}))$.
\end{proofof}

\noindent \begin{proofof}{Lem.~\ref{sec:method:theory:degree_lemma}}
	{
	\begin{displaymath}
	E[X] = \sum_{u \in V} (\frac{a_u}{b_u} \frac{b_u}{\sum_{v\in V} b_v}) = \sum_{u \in V} \frac{a_u}{\sum_{v\in V} b_v} = \frac{\sum_{u \in V} a_u}{\sum_{u \in V} b_u}.
	\end{displaymath}
	}
	As in the proof of Lemma~\ref{sec:method:theory:getnbd}, we can show $\frac{\sum_{u \in V} a_u}{\sum_{u \in V} b_u} = \frac{2|\Cm|}{2|E|} \leq 1$ and thus $E[X] \leq 1$.
\end{proofof}

\noindent \begin{proofof}{Cor.~\ref{sec:method:theory:rnd_final}}
	Due to the assumption, a node $u$ is used as the input of \getrnd with probability proportional to $deg(u)$. By Lemma~\ref{sec:method:theory:degree_lemma}, $\frac{|C^-(u)|}{deg(u)} \leq 1$ in average case. Therefore, combined with Thm.~\ref{sec:method:theory:rnd_time}, the average-case time complexity of \getrnd for $c$ samples in Alg.~\ref{algo:algo} becomes $O(c)$.
\end{proofof}

\noindent \begin{proofof}{Thm.~\ref{sec:method:theory:thm_space}}
	Alg.~\ref{algo:algo} maintains $G^{*}$ and $C$ for the current snapshot, and their size is $O(|V|+|P|+|\Cp|+|\Cm|)$ in total. 
	Moreover, additional $O(|V|)$ space is required for storing coarse cluster memberships of nodes.
	Additionally, to rapidly estimate the saving in $\varphi$, our implementation maintains the counts of edges between pairs of supernodes, and the number of nonzero counts is upper bounded by $O(|P|+|\Cp|)$. 
	Hence, the total space complexity is $O(|V|+|P|+|\Cp|+|\Cm|)$.
\end{proofof}

\vspace{-1mm}
\section{Appendix: Details of \mcmc}
\label{appendix:mcmc}
\vspace{-1mm}

\begin{algorithm}[h]
	\caption{\mcmc: a baseline algorithm \label{alg:mcmc}}
	\small
	\DontPrintSemicolon
	\KwIn{summary graph: $\GISTAR$, edge corrections: $\CI$, \\ \quad\quad\quad input change: $\{src, dst\}^+$ or $\{src, dst\}^-$}
	\KwOut{summary graph: $\GIPSTAR$, edge corrections: $\CIP$}
	\ForEach{$u$ \textup{in} \{src, dst\}}{
		$\tpsymb \gets N(u)$, $\tnsymb \gets N(u)$\;
		\ForEach{$\target$ \textup{in} $\tnsymb$}{ 
			$\cpsymb \gets V_t$\;
			darw a uniformly random node $x$ from $N(y)$\;
			draw a candidate $\candidate\in \cpsymb$ using Eq.~\eqref{appendix:proposal}\;
			temporarily move $\target$ into $S_z$\;
			compute $\Delta\varphi$ (i.e., change in $\varphi$)  for the proposal\;
			compute the acceptance probability $p$ using Eq.~\eqref{appendix:acceptance}\;
			sample $X \sim uniform(0,1)$\;
			\If{$X \leq p$}{
				accept the proposal and update $\GISTAR, \CI$\;
				\nonl \hfill $\triangleright$ see Sect.~\ref{sec:method:term} for optimal encoding
			}
		}
	}
	$\GIPSTAR, \CIP \gets \GISTAR, \CI$\;
	\Return{$\GIPSTAR$, $\CIP$}\;
\end{algorithm}

In \mcmc, described in Alg.~\ref{alg:mcmc}, we adapt the proposal probability distribution and the acceptance probability for Stochastic Block Modeling \cite{peixoto2014efficient}. 
Both are described in detail below.

\begin{figure}[h]
	\centering
	\vspace{-3mm}
	\includegraphics[width=\linewidth]{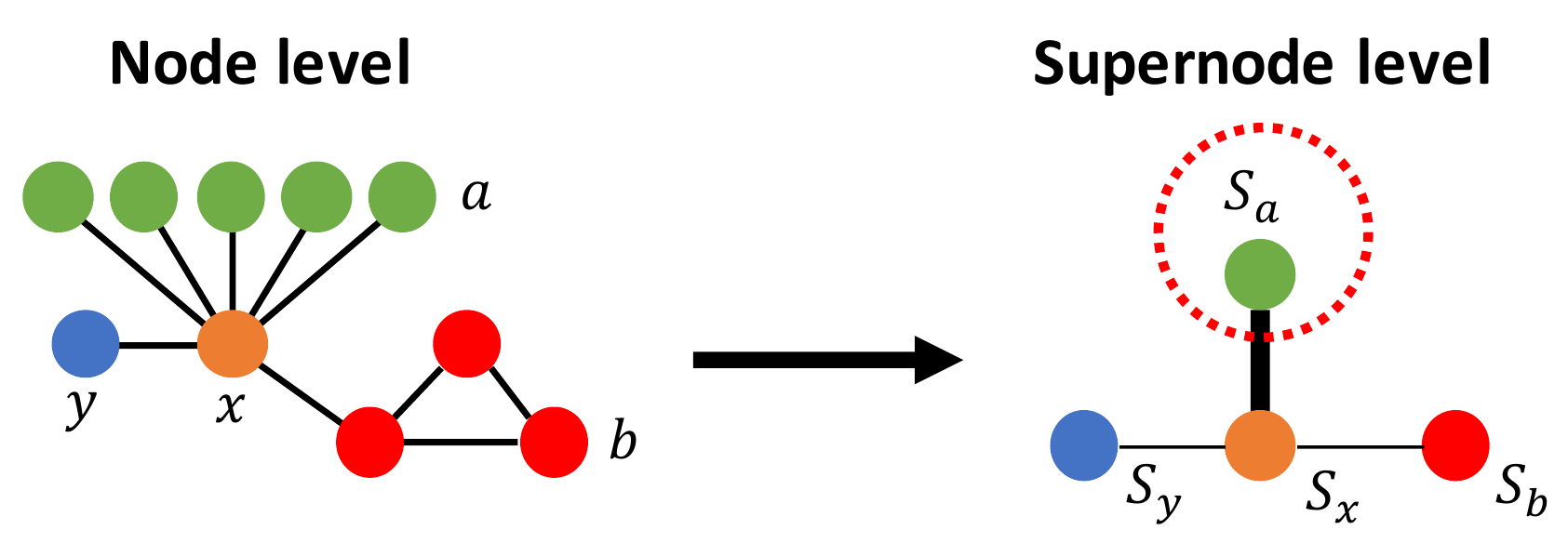} \\
	\vspace{-2mm}
	\caption{\label{fig:proposal} Proposal probability distribution in \mcmc. (Left) An input graph with the color of nodes indicating their membership to supernodes.
		(Right) Its summary graph where the thickness of each superedge indicates the number of edges between the supernodes. According to Eq.~\protect\eqref{appendix:proposal}, if $y$ is a testing node, it is most likely that one among the nodes in $S_{a}$ is selected as a candidate.}
\end{figure}

For each testing node $y$ and its uniformly random neighbor $x$, the probability that one among the nodes in each supernode $S_{\candidate}\in S$ is proposed is defined as
\begin{equation}
\label{appendix:proposal}
p(S_y \rightarrow \candidateS | S_x) := \frac{|E_{S_z,S_x}| + \epsilon}{|E_{S_x}| + \epsilon|S|},
\end{equation}
where $E_{S_x}$ is the set of edges adjacent to a node in $S_{x}$, and $\epsilon$ is a parameter that makes every supernode is selected with non-zero probability (i.e. making the process ergodic).
As illustrated in Fig.~\ref{fig:proposal}, Eq.~\eqref{appendix:proposal} places importance on supernodes densely connected by many edges to $S_x$ .

The move of $y$ into a sampled supernode $S_z$ is accepted with the  probability  defined in the Metropolis-Hastings fashion \cite{hastings1970monte} as 
\begin{equation}
\label{appendix:acceptance}
\min\left(1, e^{-\beta\Delta\varphi}\frac{\sum_{S_x\in S}p^y_{S_x} \cdot p(S_z\rightarrow S_y | S_x)}{\sum_{S_x\in S}p^y_{S_x} \cdot p(S_y\rightarrow S_z | S_x)}\right),
\end{equation}
where $p^y_{S_x} := |N(y) \cap S_x|/|N(y)|$, and $\beta$ is a parameter used to control how much randomness is injected in accepting the proposed change. The higher $\beta$ is, the more likely the algorithm is to accept the change even if the change increases $\varphi$. 
Note that $p(S_z\rightarrow S_y | S_x)$ in Eq.~\eqref{appendix:acceptance} should be computed after $y$ is moved from $S_y$ to $S_z$.

\end{document}